\documentclass[12pt]{article}
\usepackage{graphicx}
\usepackage{epstopdf}
\usepackage{subfig}
\usepackage{caption}
\usepackage{mathrsfs}
\usepackage{amssymb}
\usepackage{amsmath}
\usepackage{verbatim}
\advance\textheight by \topskip
\begin{document}


\newcommand{\HPA}[1]{{\it Helv.\ Phys.\ Acta.\ }{\bf #1}}
\newcommand{\AP}[1]{{\it Ann.\ Phys.\ }{\bf #1}}
\newcommand{\be}{\begin{equation}}
\newcommand{\ee}{\end{equation}}
\newcommand{\br}{\begin{eqnarray}}
\newcommand{\er}{\end{eqnarray}}
\newcommand{\ba}{\begin{array}}
\newcommand{\ea}{\end{array}}
\newcommand{\bi}{\begin{itemize}}
\newcommand{\ei}{\end{itemize}}
\newcommand{\bn}{\begin{enumerate}}
\newcommand{\en}{\end{enumerate}}
\newcommand{\bc}{\begin{center}}
\newcommand{\ec}{\end{center}}
\newcommand{\ul}{\underline}
\newcommand{\ol}{\overline}
\def\l{\left\langle}
\def\r{\right\rangle}
\def\as{\alpha_{s}}
\def\ycut{y_{\mbox{\tiny cut}}}
\def\yij{y_{ij}}
\def\epem{\ifmmode{e^+ e^-} \else{$e^+ e^-$} \fi}
\newcommand{\eeww}{$e^+e^-\rightarrow W^+ W^-$}
\newcommand{\qqQQ}{$q_1\bar q_2 Q_3\bar Q_4$}
\newcommand{\eeqqQQ}{$e^+e^-\rightarrow q_1\bar q_2 Q_3\bar Q_4$}
\newcommand{\eewwqqqq}{$e^+e^-\rightarrow W^+ W^-\ar q\bar q Q\bar Q$}
\newcommand{\eeqqgg}{$e^+e^-\rightarrow q\bar q gg$}
\newcommand{\eeqloop}{$e^+e^-\rightarrow q\bar q gg$ via loop of quarks}
\newcommand{\eeqqqq}{$e^+e^-\rightarrow q\bar q Q\bar Q$}
\newcommand{\eewwjjjj}{$e^+e^-\rightarrow W^+ W^-\rightarrow 4~{\rm{jet}}$}
\newcommand{\eeqqggjjjj}{$e^+e^-\rightarrow q\bar 
q gg\rightarrow 4~{\rm{jet}}$}
\newcommand{\eeqloopjjjj}{$e^+e^-\rightarrow q\bar 
q gg\rightarrow 4~{\rm{jet}}$ via loop of quarks}
\newcommand{\eeqqqqjjjj}{$e^+e^-\rightarrow q\bar q Q\bar Q\rightarrow
4~{\rm{jet}}$}
\newcommand{\eejjjj}{$e^+e^-\rightarrow 4~{\rm{jet}}$}
\newcommand{\jjjj}{$4~{\rm{jet}}$}
\newcommand{\qqbar}{$q\bar q$}
\newcommand{\ww}{$W^+W^-$}
\newcommand{\ar}{\rightarrow}
\newcommand{\sm}{${\cal {SM}}$}
\newcommand{\Dir}{\kern -6.4pt\Big{/}}
\newcommand{\Dirin}{\kern -10.4pt\Big{/}\kern 4.4pt}
\newcommand{\DDir}{\kern -8.0pt\Big{/}}
\newcommand{\DGir}{\kern -6.0pt\Big{/}}
\newcommand{\wwqqqq}{$W^+ W^-\ar q\bar q Q\bar Q$}
\newcommand{\qqgg}{$q\bar q gg$}
\newcommand{\qloop}{$q\bar q gg$ via loop of quarks}
\newcommand{\qqqq}{$q\bar q Q\bar Q$}

\def\st{\sigma_{\mbox{\scriptsize t}}}
\def\Ord{\buildrel{\scriptscriptstyle <}\over{\scriptscriptstyle\sim}}
\def\OOrd{\buildrel{\scriptscriptstyle >}\over{\scriptscriptstyle\sim}}
\def\jhep #1 #2 #3 {{JHEP} {\bf#1} (#2) #3}
\def\plb #1 #2 #3 {{Phys.~Lett.} {\bf B#1} (#2) #3}
\def\npb #1 #2 #3 {{Nucl.~Phys.} {\bf B#1} (#2) #3}
\def\epjc #1 #2 #3 {{Eur.~Phys.~J.} {\bf C#1} (#2) #3}
\def\zpc #1 #2 #3 {{Z.~Phys.} {\bf C#1} (#2) #3}
\def\jpg #1 #2 #3 {{J.~Phys.} {\bf G#1} (#2) #3}
\def\prd #1 #2 #3 {{Phys.~Rev.} {\bf D#1} (#2) #3}
\def\prep #1 #2 #3 {{Phys.~Rep.} {\bf#1} (#2) #3}
\def\prl #1 #2 #3 {{Phys.~Rev.~Lett.} {\bf#1} (#2) #3}
\def\mpl #1 #2 #3 {{Mod.~Phys.~Lett.} {\bf#1} (#2) #3}
\def\rmp #1 #2 #3 {{Rev. Mod. Phys.} {\bf#1} (#2) #3}
\def\cpc #1 #2 #3 {{Comp. Phys. Commun.} {\bf#1} (#2) #3}
\def\sjnp #1 #2 #3 {{Sov. J. Nucl. Phys.} {\bf#1} (#2) #3}
\def\xx #1 #2 #3 {{\bf#1}, (#2) #3}
\def\hepph #1 {{\tt hep-ph/#1}}
\def\preprint{{preprint}}

\def\beq{\begin{equation}}
\def\beeq{\begin{eqnarray}}
\def\eeq{\end{equation}}
\def\eeeq{\end{eqnarray}}
\def\a0{\bar\alpha_0}
\def\thrust{\mbox{T}}
\def\Thrust{\mathrm{\tiny T}}
\def\ae{\alpha_{\mbox{\scriptsize eff}}}
\def\ap{\bar\alpha_p}
\def\as{\alpha_{\mathrm{S}}}
\def\aem{\alpha_{\mathrm{EM}}}
\def\b0{\beta_0}
\def\cN{{\cal N}}
\def\cd{\chi^2/\mbox{d.o.f.}}
\def\Ecm{E_{\mbox{\scriptsize cm}}}
\def\ee{e^+e^-}
\def\enap{\mbox{e}}
\def\eps{\epsilon}
\def\ex{{\mbox{\scriptsize exp}}}
\def\GeV{\mbox{\rm{GeV}}}
\def\half{{\textstyle {1\over2}}}
\def\jet{{\mbox{\scriptsize jet}}}
\def\kij{k^2_{\bot ij}}
\def\kp{k_\perp}
\def\kps{k_\perp^2}
\def\kt{k_\bot}
\def\lms{\Lambda^{(n_{\rm f}=4)}_{\overline{\mathrm{MS}}}}
\def\mI{\mu_{\mathrm{I}}}
\def\mR{\mu_{\mathrm{R}}}
\def\MSbar{\overline{\mathrm{MS}}}
\def\mx{{\mbox{\scriptsize max}}}
\def\NP{{\mathrm{NP}}}
\def\pd{\partial}
\def\pt{{\mbox{\scriptsize pert}}}
\def\pw{{\mbox{\scriptsize pow}}}
\def\so{{\mbox{\scriptsize soft}}}
\def\st{\sigma_{\mbox{\scriptsize tot}}}
\def\ycut{y_{\mathrm{cut}}}
\def\slashchar#1{\setbox0=\hbox{$#1$}           
     \dimen0=\wd0                                 
     \setbox1=\hbox{/} \dimen1=\wd1               
     \ifdim\dimen0>\dimen1                        
        \rlap{\hbox to \dimen0{\hfil/\hfil}}      
        #1                                        
     \else                                        
        \rlap{\hbox to \dimen1{\hfil$#1$\hfil}}   
        /                                         
     \fi}                                         %
\def\etmiss{\slashchar{E}^T}
\def\Meff{M_{\rm eff}}
\def\Ord{\lsim}
\def\OOrd{\gsim}
\def\tq{\tilde q}
\def\tchi{\tilde\chi}
\def\lsp{\tilde\chi_1^0}

\def\gam{\gamma}
\def\ph{\gamma}
\def\be{\begin{equation}}
\def\ee{\end{equation}}
\def\bea{\begin{eqnarray}}
\def\eea{\end{eqnarray}}
\def\lsim{\:\raisebox{-0.5ex}{$\stackrel{\textstyle<}{\sim}$}\:}
\def\gsim{\:\raisebox{-0.5ex}{$\stackrel{\textstyle>}{\sim}$}\:}

\def\ino{\mathaccent"7E} \def\gluino{\ino{g}} \def\mgluino{m_{\gluino}}
\def\sqk{\ino{q}} \def\sup{\ino{u}} \def\sdn{\ino{d}}
\def\chargino{\ino{\omega}} \def\neutralino{\ino{\chi}}
\def\cab{\ensuremath{C_{\alpha\beta}}} \def\proj{\ensuremath{\mathcal P}}
\def\dab{\delta_{\alpha\beta}}
\def\zz{s-M_Z^2+iM_Z\Gamma_Z} \def\zw{s-M_W^2+iM_W\Gamma_W}
\def\prop{\ensuremath{\mathcal G}} \def\ckm{\ensuremath{V_{\rm CKM}^2}}
\def\aem{\alpha_{\rm EM}} \def\stw{s_{2W}} \def\sttw{s_{2W}^2}
\def\nc{N_C} \def\cf{C_F} \def\ca{C_A}
\def\qcd{\textsc{Qcd}} \def\susy{supersymmetric} \def\mssm{\textsc{Mssm}}
\def\slash{/\kern -5pt} \def\stick{\rule[-0.2cm]{0cm}{0.6cm}}
\def\h{\hspace*{-0.3cm}}

\def\ims #1 {\ensuremath{M^2_{[#1]}}}
\def\tw{\tilde \chi^\pm}
\def\tz{\tilde \chi^0}
\def\tf{\tilde f}
\def\tl{\tilde l}
\def\ppb{p\bar{p}}
\def\gl{\tilde{g}}
\def\sq{\tilde{q}}
\def\sqb{{\tilde{q}}^*}
\def\qb{\bar{q}}
\def\sqL{\tilde{q}_{_L}}
\def\sqR{\tilde{q}_{_R}}
\def\ms{m_{\tilde q}}
\def\mg{m_{\tilde g}}
\def\Gs{\Gamma_{\tilde q}}
\def\Gg{\Gamma_{\tilde g}}
\def\md{m_{-}}
\def\eps{\varepsilon}
\def\Ce{C_\eps}
\def\dnq{\frac{d^nq}{(2\pi)^n}}
\def\DR{$\overline{DR}$\,\,}
\def\MS{$\overline{MS}$\,\,}
\def\DRm{\overline{DR}}
\def\MSm{\overline{MS}}
\def\ghat{\hat{g}_s}
\def\shat{\hat{s}}
\def\sihat{\hat{\sigma}}
\def\Li{\text{Li}_2}
\def\bs{\beta_{\sq}}
\def\xs{x_{\sq}}
\def\xsa{x_{1\sq}}
\def\xsb{x_{2\sq}}
\def\bg{\beta_{\gl}}
\def\xg{x_{\gl}}
\def\xga{x_{1\gl}}
\def\xgb{x_{2\gl}}
\def\lsp{\tilde{\chi}_1^0}

\def\gluino{\mathaccent"7E g}
\def\mgluino{m_{\gluino}}
\def\squark{\mathaccent"7E q}
\def\msquark{m_{\mathaccent"7E q}}
\def\M{ \overline{|\mathcal{M}|^2} }
\def\utm{ut-M_a^2M_b^2}
\def\MiLR{M_{i_{L,R}}}
\def\MiRL{M_{i_{R,L}}}
\def\MjLR{M_{j_{L,R}}}
\def\MjRL{M_{j_{R,L}}}
\def\tiLR{t_{i_{L,R}}}
\def\tiRL{t_{i_{R,L}}}
\def\tjLR{t_{j_{L,R}}}
\def\tjRL{t_{j_{R,L}}}
\def\tg{t_{\gluino}}
\def\uiLR{u_{i_{L,R}}}
\def\uiRL{u_{i_{R,L}}}
\def\ujLR{u_{j_{L,R}}}
\def\ujRL{u_{j_{R,L}}}
\def\ug{u_{\gluino}}
\def\utot{u \leftrightarrow t}
\def\ar{\to}
\def\sqk{\mathaccent"7E q}
\def\sup{\mathaccent"7E u}
\def\sdn{\mathaccent"7E d}
\def\chargino{\mathaccent"7E \chi}
\def\neutralino{\mathaccent"7E \chi}
\def\slepton{\mathaccent"7E l}
\def\M{ \overline{|\mathcal{M}|^2} }
\def\cab{\ensuremath{C_{\alpha\beta}}}
\def\ckm{\ensuremath{V_{\rm CKM}^2}}
\def\zz{s-M_Z^2+iM_Z\Gamma_Z}
\def\zw{s-M_W^2+iM_W\Gamma_W}
\def\s22w{s_{2W}^2}

\newcommand{\cpmtwo}    {\mbox{$ {\chi}^{\pm}_{2}                    $}}
\newcommand{\cpmone}    {\mbox{$ {\chi}^{\pm}_{1}                    $}}

\begin{flushright}
{SHEP-09-05}\\
\today
\end{flushright}
\vskip0.1cm\noindent
\begin{center}
{{\Large {\bf Probing the $Z'$ sector of the minimal $B-L$ model\\[0.25cm]
at future Linear Colliders in the
$e^+e^-\to \mu^+\mu^-$ process}}
\\[1.0cm]
{\large L. Basso$^1$, A. Belyaev$^1$, S. Moretti$^{1,2}$ and G. M. Pruna$^1$}\\[0.30 cm]
{\it  $^1$School of Physics and Astronomy, University of Southampton,}\\
{\it  Highfield, Southampton SO17 1BJ, UK.}\\
{\it  $^2$Dipartimento di Fisica Teorica, Universit\`a di Torino,}\\
{\it  Via Pietro Giuria 1, 10125 Torino, Italy.}
}
\\[1.25cm]
\end{center}

\begin{abstract}
{\small
\noindent
We study the capabilities of future electron-positron  Linear Colliders, with
centre-of-mass {energy} at the TeV scale, in accessing the parameter space
of   {a $Z'$ boson within  the minimal $B-L$ model}.  In such a model, wherein
the Standard Model gauge group is augmented by a broken $U(1)_{B-L}$ symmetry
-- with $B(L)$ being the baryon(lepton) number -- the emerging  $Z'$ mass is
expected to be in the above energy range. 
We carry out a detailed comparison between the discovery regions mapped
over a two-dimensional configuration space ($Z'$ mass and coupling) at
the Large Hadron Collider and possible future Linear Colliders 
for the case of di-muon production. As known in the literature
for other $Z'$ models, we confirm that  leptonic machines, 
as compared to the CERN hadronic accelerator,
display {an additional} potential in discovering a $B-L$
$Z'$ boson as well
as in allowing one to study its properties at a level of
precision well beyond that of {any of the existing colliders.}
}
\end{abstract}


\section{Introduction}
\label{Sec:Intro}

The $B-L$ (baryon number minus lepton number) symmetry plays an
important role in various physics scenarios beyond the Standard Model
(SM). Firstly, the gauged  $U(1)_{B-L}$ symmetry group is  contained
in a  Grand Unified Theory (GUT) described by a $SO(10)$
group~\cite{Buchmuller:1991ce}.  Secondly, the scale of  the $B-L$
symmetry  breaking is related to the mass scale of the heavy
right-handed Majorana neutrino mass terms providing the well-known
see-saw mechanism~\cite{see-saw} of light neutrino mass
generation. Thirdly, the $B-L$ symmetry and the scale of its breaking
are tightly connected to the baryogenesis mechanism through
leptogenesis~\cite{Fukugita:1986hr} via sphaleron interactions
preserving $B-L$.

The minimal $B-L$ low-energy extension of the SM consists of a further
$U(1)_{B-L}$ gauge group, three right-handed neutrinos and an
additional Higgs boson generated through the $U(1)_{B-L}$ symmetry
breaking. It is important to note that in this model the ${B-L}$
breaking can take place at the TeV scale, i.e. far below that of any
GUT. This $B-L$ scenario therefore has interesting implications at the
Large Hadron Collider (LHC), including  new clean signatures from  
$Z'$, Higgs bosons and heavy neutrinos~\cite{B-L:LHC}--\cite{B-L:rev}.

In the present paper we study the phenomenology related to the $Z'$
sector of the minimal $B-L$  extension of the SM at the new generation
of $e^+ e^-$ Linear Colliders (LCs) \cite{LCs}. We consider the
$e^+e^-\to \mu^+\mu^-$ channel as a representative process in order to
study new signatures pertaining to the $B-L$ model.

As it is well known (see, e.g., Refs.~\cite{rizzo} and \cite{LCZZ}), the 
LC environment
is one of the most suitable for $Z'$ physics, for two main reasons.  
First, if a $Z'$ is found at the LHC, it could be the case that the
underlying model is hard to identify at the
hadronic machine; in contrast, the clean experimental environment of a
LC is the ideal framework to establish  the $Z'$ line-shape (i.e. its
mass and width) and to measure its couplings, thereby 
identifying the model and the observed spin$-1$
boson~\cite{ILC_RDR}. Second, we will also show that there exists
further scope for a LC operating at TeV energies: specifically, to
discover a $Z'$ boson over regions of the $B-L$
parameter space which
cannot be probed at all at the LHC, either directly through a
resonance (when $\sqrt s_{e^+e^-}\geq M_{Z'}$) or indirectly through
interference effects (when $\sqrt s_{e^+e^-}< M_{Z'}$). In both
instances, a LC proves to be more powerful than the LHC in accessing
the region of small $Z'$ couplings.

This work is organised as follows. In the next section we describe
the model. In Sect.~\ref{Sec:Calculation} we illustrate the
computational techniques adopted. 
In Sect.~\ref{Sec:Results} we present
our numerical results. The conclusions are in Sect.~\ref{Sec:Conclusions}.


\section{The model}
\label{Sec:Model}
The model under study is the so-called ``pure'' or ``minimal''
$B-L$ model (see \cite{B-L:LHC} for conventions and references) 
since it has vanishing mixing between the two $U(1)_{Y}$ 
and $U(1)_{B-L}$ groups.
In the rest of this paper we refer to this model simply as the ``$B-L$
model''.  {In this model the} classical gauge invariant Lagrangian,
obeying the $SU(3)_C\times SU(2)_L\times U(1)_Y\times U(1)_{B-L}$
gauge symmetry, can be decomposed as:
\begin{equation}\label{L}
\mathscr{L}=\mathscr{L}_{YM} + \mathscr{L}_s + \mathscr{L}_f + \mathscr{L}_Y \, .
\end{equation}
The non-Abelian field strengths in $\mathscr{L}_{YM}$ are the same as in the SM
whereas the Abelian
ones can be written as follows:
\begin{equation}\label{La}
\mathscr{L}^{\rm Abel}_{YM} = 
-\frac{1}{4}F^{\mu\nu}F_{\mu\nu}-\frac{1}{4}F^{\prime\mu\nu}F^\prime _{\mu\nu}\, ,
\end{equation}
where
\begin{eqnarray}\label{new-fs3}
F_{\mu\nu}		&=&	\partial _{\mu}B_{\nu} - \partial _{\nu}B_{\mu} \, , \\ \label{new-fs4}
F^\prime_{\mu\nu}	&=&	\partial _{\mu}B^\prime_{\nu} - \partial _{\nu}B^\prime_{\mu} \, .
\end{eqnarray}
In this field basis, the covariant derivative is:
\begin{equation}\label{cov_der}
D_{\mu}\equiv \partial _{\mu} + ig_S T^{\alpha}G_{\mu}^{\phantom{o}\alpha} 
+ igT^aW_{\mu}^{\phantom{o}a} +ig_1YB_{\mu} +i(\widetilde{g}Y + g_1'Y_{B-L})B'_{\mu}\, .
\end{equation}
The ``pure'' or ``minimal'' $B-L$ model is defined by the condition $\widetilde{g} = 0$, that implies no mixing between the $Z'$ and the SM-$Z$ gauge bosons.

The fermionic Lagrangian (where $k$ is the
generation index) is given by
\begin{eqnarray} \nonumber
\mathscr{L}_f &=& \sum _{k=1}^3 \Big( i\overline {q_{kL}} \gamma _{\mu}D^{\mu} q_{kL} + i\overline {u_{kR}}
			\gamma _{\mu}D^{\mu} u_{kR} +i\overline {d_{kR}} \gamma _{\mu}D^{\mu} d_{kR} +\\
			  && + i\overline {l_{kL}} \gamma _{\mu}D^{\mu} l_{kL} + i\overline {e_{kR}}
			\gamma _{\mu}D^{\mu} e_{kR} +i\overline {\nu _{kR}} \gamma _{\mu}D^{\mu} \nu
			_{kR} \Big)  \, ,
\end{eqnarray}
 where the fields' charges are the usual SM and $B-L$ ones (in particular, $B-L = 1/3$ for quarks and $-1$ for leptons).
  The  $B-L$ charge assignments of the fields
  as well as the introduction of new
  fermionic  right-handed heavy neutrinos ($\nu_R$) and
  scalar Higgs ($\chi$, charged $+2$ under $B-L$)  
  fields are designed to eliminate the triangular $B-L$  gauge anomalies and to ensure the gauge invariance of the theory (see eq. (\ref{L_Yukawa})), respectively.
  Therefore, the $B-L$  gauge extension of the SM group
  broken at the EW scale does necessarily require
  at least one new scalar field and three new fermionic fields which are
  charged with respect to the $B-L$ group.

The scalar Lagrangian is:
\begin{equation}\label{new-scalar_L}
\mathscr{L}_s=\left( D^{\mu} H\right) ^{\dagger} D_{\mu}H + 
\left( D^{\mu} \chi\right) ^{\dagger} D_{\mu}\chi - V(H,\chi ) \, ,
\end{equation}
{with the scalar potential given by}
\begin{equation}\label{new-potential}
V(H,\chi ) = m^2H^{\dagger}H +
 \mu ^2\mid\chi\mid ^2 +
  \lambda _1 (H^{\dagger}H)^2 +\lambda _2 \mid\chi\mid ^4 + \lambda _3 H^{\dagger}H\mid\chi\mid ^2  \, ,
\end{equation}
{where $H$ and $\chi$ are the complex scalar Higgs 
doublet and singlet fields, respectively.}

Finally, the Yukawa interactions are:
\begin{eqnarray}\nonumber
\mathscr{L}_Y &=& -y^d_{jk}\overline {q_{jL}} d_{kR}H 
                 -y^u_{jk}\overline {q_{jL}} u_{kR}\widetilde H 
		 -y^e_{jk}\overline {l_{jL}} e_{kR}H \\ \label{L_Yukawa}
	      & & -y^{\nu}_{jk}\overline {l_{jL}} \nu _{kR}\widetilde H 
	         -y^M_{jk}\overline {(\nu _R)^c_j} \nu _{kR}\chi +  {\rm 
h.c.}  \, ,
\end{eqnarray}
{where $\tilde H=i\sigma^2 H^*$ and  $i,j,k$ take the values $1$ to $3$},
where the last term is the Majorana contribution
 and the others the usual Dirac 
ones.


\section{Calculation}
\label{Sec:Calculation}

The study we present in this paper has been performed  with the 
help of the CalcHEP package \cite{calchep}, in which the model 
under discussion had been previously implemented via the 
LanHEP tool \cite{lanhep}, 
as already discussed in \cite{B-L:LHC}. 

A feature specific to LCs is the presence of Initial State Radiation (ISR) and Beamstrahlung. For the former, 
CalcHEP \cite{calchep_man} implements the Jadach, Skrzypek and Ward expressions of Ref.~\cite{ISR}. 
Regarding the latter, we adopted the parameterisation specified for the 
International Linear Collider
(ILC) project in \cite{ILC_RDR}:
\begin{eqnarray}
\mbox{Horizontal beam  size (nm)} &=& 640,   \nonumber \\
\mbox{Vertical   beam  size (nm)} &=& 5.7,   \nonumber \\
\mbox{Bunch length (mm)}          &=& 0.300, \nonumber \\
\mbox{Number of particles in the bunch (N)} &=& 2\times10^{10}.
\end{eqnarray}

There exists a certain  subtlety in  the comparison of the  LHC and LC
discovery potentials of a  $Z'$ boson. This comparison is not
straightforward and ought to be performed
carefully~\cite{AguilarSaavedra:2005pw}--\cite{Weiglein:2004hn}.
First of all, we need to compare consistent temporal collections of data.
On the one hand,
luminosities are different at the two kind of machines and so are
{supposed to be the} running
schedules. Besides, 
{in this work, we also consider the fact that, while at the LHC we will
have essentially
a fixed beam energy technology, at LCs one can afford the
possibility of beam energy scans.}
In this connection, while comparing the scope of the two,
we have assumed 100 fb$^{-1}$ for the LHC throughout and 500(10) fb$^{-1}$ 
for LCs running at fixed energy (in energy scanning mode). 
On the other hand, data
samples will be collected differently, chiefly, acceptance and selection
 procedures will be different. In this connection, we have assumed standard
acceptance cuts (on muons) at the LHC and a typical LC\footnote{These cuts
will then only be applied in the case of Figs.~\ref{contour1}
and \ref{contour2} (i.e., combination of eqs. (\ref{LHC_cut}) and (\ref{LHC_ris}) for the LHC whereas eqs. (\ref{LC_cut}) and (\ref{LC_ris}) for a LC)  
and of Fig.~\ref{LHC_reach_3TeV} (again, combination of eqs. (\ref{LHC_cut}) and (\ref{LHC_ris}) for the LHC)
and not elsewhere.},
\begin{eqnarray}\label{LHC_cut}
{\rm LHC:}\qquad p_T^\mu&>& 10~{\rm GeV},\qquad |\eta^\mu|<2.5,\\ \label{LC_cut}
{\rm LC:}\qquad     E^\mu &>& 10~{\rm GeV},\qquad |\cos\theta^\mu| < 0.95.
\end{eqnarray}

Then,
for both signal and  background, we apply 
the following cut on the di-muon invariant mass,  $M_{\mu\mu}$:
\begin{eqnarray}\label{LHC_ris}
{\rm LHC:}\qquad |M_{\mu\mu}-M_{Z'}|&<& 
\mbox{max} \left( 3\Gamma_{Z'},\; \left( 0.03\sqrt{\frac{M_{Z'}}{\rm GeV}}
+0.005\frac{M_{Z'}}{\rm GeV} \right) {\rm GeV}\; \right), \\ \label{LC_ris}
{\rm LC:}\qquad |M_{\mu\mu}-M_{Z'}|&<& 
\mbox{max} \left( 3\Gamma_{Z'},\; 0.15\sqrt{\frac{M_{Z'}}{\rm GeV}}
{\rm GeV}\; \right),
\end{eqnarray}
that is, a half 
window as large as either three times the width of the $Z'$-boson
or the di-muon mass resolution\footnote{We assume  the CMS di-muon mass
resolution \cite{CMSdet} for the LHC environment and the ILC prototype
di-muon mass resolution \cite{ILCdet} for typical LCs detectors.}, 
whichever the largest.

{
In our analysis we implement a suitable definition of signal significance,
applicable to both the LHC and LC contexts, which we have done as follows. 
In the region where the number
of both signal ($s$) and background ($b$) 
events is large enough (bigger than 20),
we use a definition of significance
based on Gaussian statistics, ${\sigma} \equiv {\it s}/{\sqrt{b}}$.
Otherwise, in case of lower statistics,
we exploited the 
Bityukov algorithm \cite{Bityukov}, which basically uses the 
Poisson `true' distribution instead of the approximated Gaussian one. 
}
Hereafter, to `Observation' it will correspond the condition
$\sigma\geq3$ and to `Discovery' $\sigma\geq5$.

Finally, as in \cite{B-L:LHC}, in the LHC case we used CTEQ6L \cite{CTEQ},  
with $Q^2=M_{Z'}^2$, as default Parton Distribution Functions (PDFs).\\


\section{Results}
\label{Sec:Results}
Hereafter,
we
assume that the heavy neutrinos and Higgs states of the model have masses
as in \cite{B-L:LHC}\footnote{For sake of  completeness, we state here again
the values we chose in \cite{B-L:LHC}: $m_{\nu ^1_h} = m_{\nu ^2_h} =
m_{\nu ^3_h} = 200$ GeV and $m_{h_1} = 125$ GeV, $m_{h_2} = 450$ GeV, for
the heavy neutrino and Higgs masses, respectively.}. This
choice of the parameters only affects the $Z'$ width, in fact minimally
(a few percents), so that our conclusions will be unchanged by it.
{
Regarding the possible phenomenology of the new neutrino states, the relatively
small cross sections involving the production of the latter require very high
luminosity to become important, especially for very small values of
the couplings, hence beyond the scope of the present
paper\footnote{The phenomenology of our $Z'$ involving the new heavy
neutrinos has been developed in the LHC framework in \cite{B-L:LHC}: we
remand to it for further details.}. Concerning the Higgs sector, we
are currently in the process of defining the accessible parameter space
(subject to experimental and theoretical constraints) ameanable to
phenomenological analysis \cite{preparation}. The Higgs mass choices made
here are then meant to be illustrative of the case in which the
Higgs sector of the model impinges
marginally on $Z'$ phenomenology. 
}

\subsection{Experimental limits on $Z'$ masses and couplings in $B-L$}\label{Exp_lim}
Before proceeding to our signal-to-background analysis, we ought to
define the  parameter space of the $B-L$ model 
sector, compliant with current experimental constraints. Some stringent
`indirect' limits  on the $Z'$ mass-to-coupling ratio can be extracted
from precision data (obtained at LEP and SLC), where the use of a
four-fermion interaction 
already gives rather
accurate results \cite{Carena}. Despite this approach is well
established, it is worth to note that more sophisticated techniques 
could change such bounds\footnote{For example, like those in
Ref.~\cite{Cacciapaglia:2006pk}, based on an effective Lagrangian
parameterisation.
}. 
However, in the course of our analysis, we will be constraining ourselves
to regions of masses and couplings that are immune from such constraints,
as they lie well beyond the LEP and SLC limits  (as well illustrated in
some of our plots). Since the approximation 
used for the
extraction of such limits is therefore irrelevant, we decided to quote
and adopt here the more conservative result obtained by
\cite{Cacciapaglia:2006pk}:
\begin{equation}\label{LEP_bound}
\frac{M_{Z'}}{g'_1} \geq 7\; \rm{TeV}\, 
\end{equation}
(which is not significantly
lowered in the analysis of 
\cite{Carena}: where
$M_{Z'}/g'_1 \geq 6$ TeV is quoted). The most constraining `direct'
bounds come from Run 2 at Tevatron, chiefly from  $q\overline{q}
\rightarrow \mu^+ \mu^-$ analyses. For definiteness, we take the  CDF
analysis of Ref.~\cite{Tevatron_2.3fb} using $2.3\,\mbox{fb}^{-1}$ of
data, which sets lower limits for $Z'$ masses coming from several
scenarios (e.g., a SM-like $Z'$ and some $E_6$ string-inspired $Z'$
models),  but not for the $B-L$ case. 
Nonetheless, by rescaling the SM-like $Z'$ coupling, we get for our
$B-L$ setup, at $95\%$ C.L.{, the lower bounds displayed in Tab.~\ref{mzp-low_bound}.}
\begin{table}[h]
\begin{center}
\begin{tabular}{|c|c|}
\hline
 $g_1'$         & $M_{Z'}$ (GeV)\\
\hline
 0.065		& 600        \\  
 0.075		& 680        \\ 
 0.090		& 740        \\ 
 0.1		& 800        \\ 
 0.2		& 960        \\
 0.5		& 1140       \\
\hline
\end{tabular}
\end{center}
\vskip -0.5cm
\caption{Lower bounds on the $Z'$ mass for selected $g_1'$ values in our 
$B-L$ model, at $95\%$ C.L.,
by rescaling the SM-like $Z'$ coupling of Ref.~\cite{Tevatron_2.3fb}. 
\label{mzp-low_bound}}
\end{table}

\subsection{The LHC and LC potential in detecting $Z'$ bosons in $B-L$ }
\begin{figure}[h]
  \subfloat[]{ 
  \label{contour1}
  \includegraphics[angle=0,width=0.48\textwidth ]{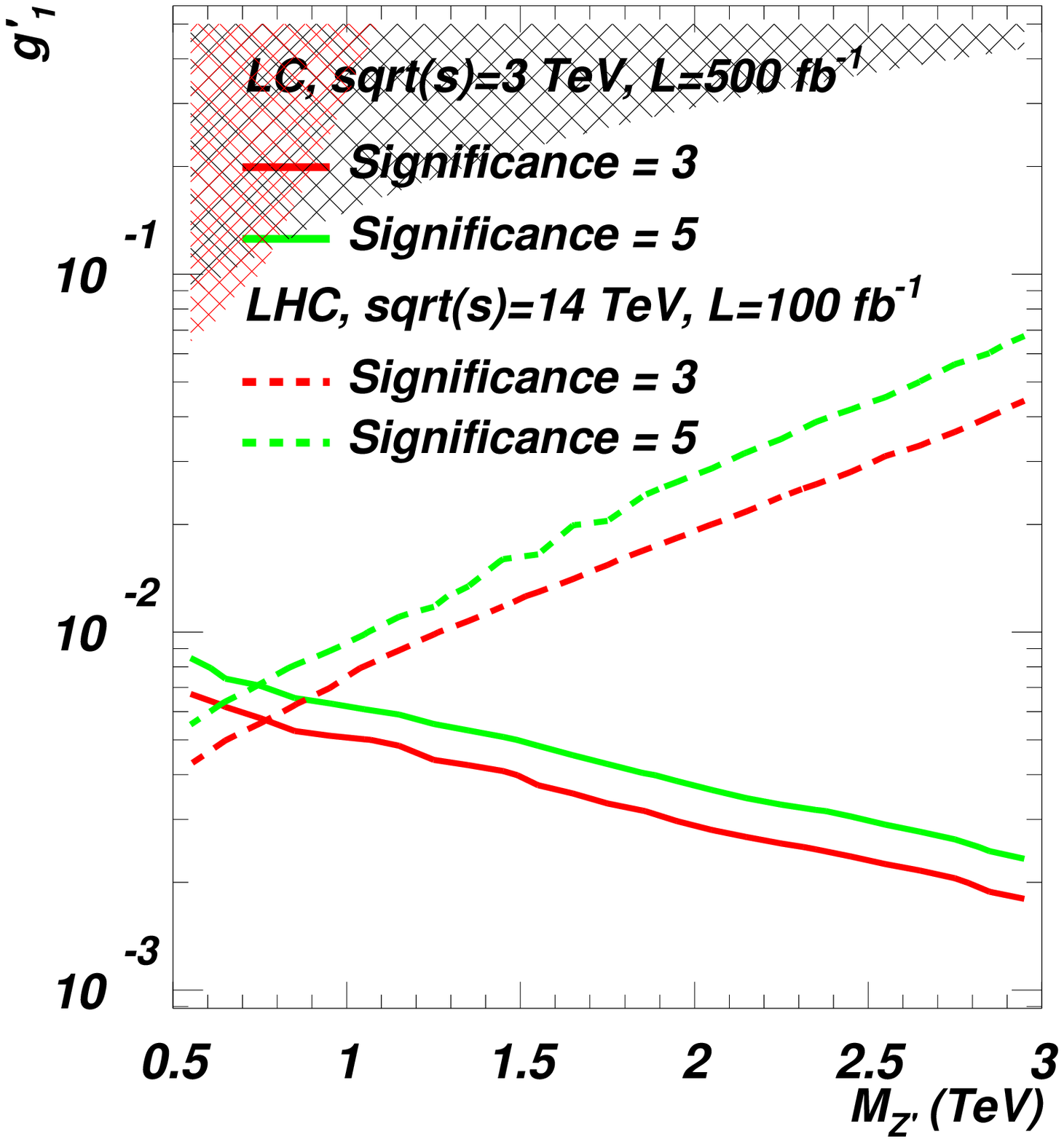}}
  \subfloat[]{
  \label{contour2}
  \includegraphics[angle=0,width=0.48\textwidth ]{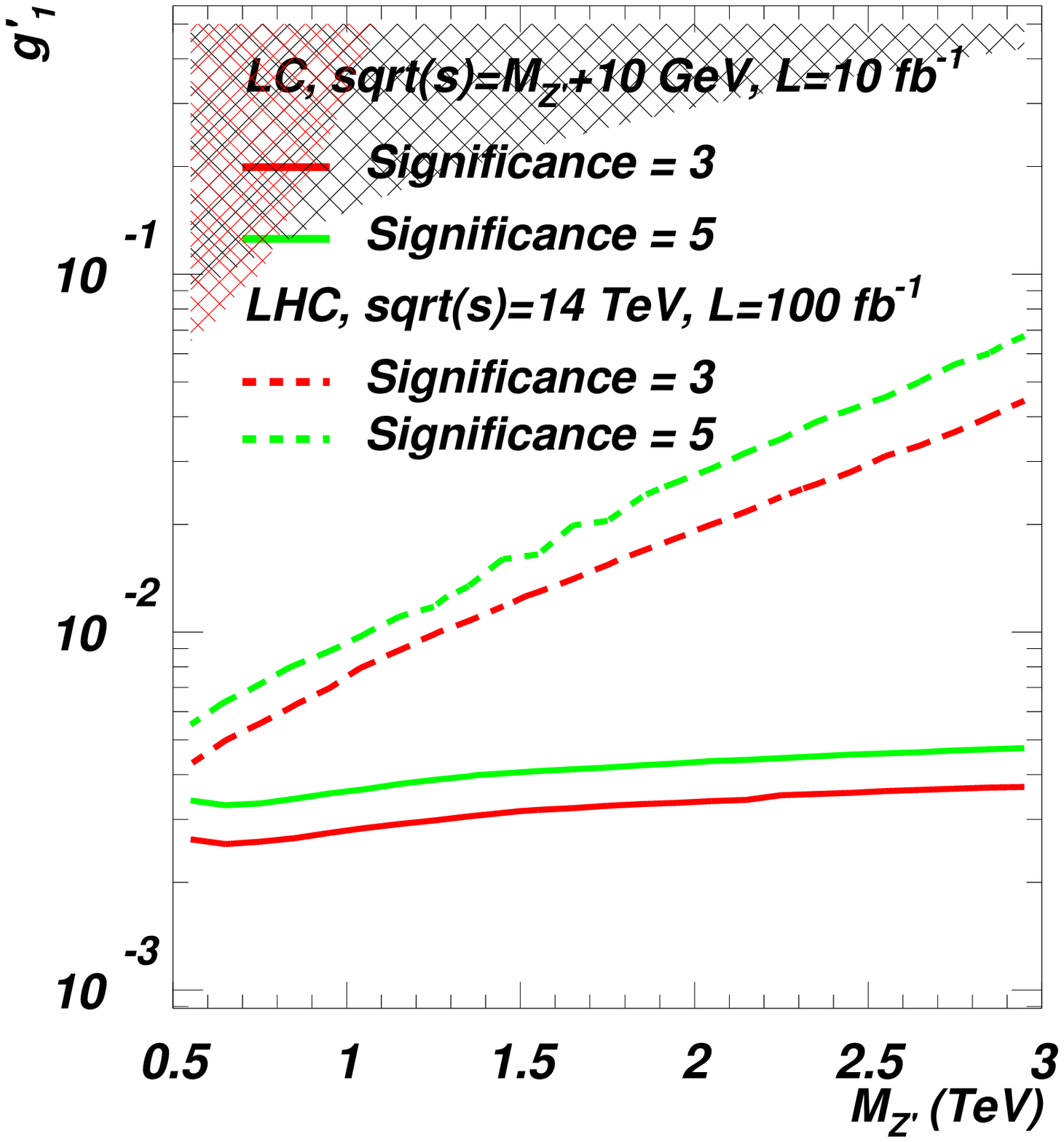}}
  \vspace*{-0.5cm}
  \caption{Significance contour levels plotted against $g_1'$
and $M_{Z'}$ both at the LHC for $L=100\;{\rm fb}^{-1}$
($\sqrt{s_{pp}}=14$ TeV, dotted line) and (\ref{contour1}) a LC for 
$L=500\;{\rm fb}^{-1}$, $\sqrt{s_{e^+e^-}}=3$ TeV plus (\ref{contour2})
a LC for 
$L=10\;{\rm fb}^{-1}$, $\sqrt{s_{e^+e^-}}=M_{Z'}+10$ GeV, both in continuous line. The shaded areas correspond to the region of parameter space excluded
experimentally, in accordance with eq.~(\ref{LEP_bound}) (LEP bounds, in black) and Tab.~\ref{mzp-low_bound} (Tevatron bounds, in red).}
  \label{Disc_power}
\end{figure}

%
%
We start the presentation of our results by showing Fig.~\ref{Disc_power},
which demonstrates the LHC and ILC discovery potential of a $Z'$ boson
over the $M_{Z'}$-$g_1'$ plane.
{Here, we define the signal as 
di-muon production via $Z'$ exchange together with
its interferences with the SM (i.e., $\gamma$ and $Z$ exchange) sub-processes
whereas as background we take the  SM
di-muon production via  $\gamma$ and $Z$ exchange}. 
Both signal and background are then limited to the
detector acceptance volumes and $M_{\mu\mu}$ invariant mass window described in
the previous section.
In Fig.~\ref{contour1} we considered a LC collecting data at the fixed
energy of $\sqrt{s_{e^+e^-}}=3$ TeV.
As one can clearly see,
for $M_{Z'}>800$ GeV, the LC potential to explore the $M_{Z'}$-$g_1'$
parameter space goes beyond the LHC reach.
For example, for $M_{Z'}=1$ TeV, the
LHC can discover a $Z'$ if $g'\approx 0.007$
while a LC can achieve this for $g'\approx 0.005$.
The difference is even  more drastic for larger $Z'$ masses
as one can see from Tab.~\ref{mzp-gp-tab}:
 a LC can discover a $Z'$ with a $2$ TeV mass for a 
$g_1'$ coupling which is a factor 8 smaller than the one
for which the same mass $Z'$ can be discovered at the LHC.
\begin{table}[h]
\begin{center}
\begin{tabular}{|l|l|l|l|}
\hline
$M_{Z'}$ (TeV)  & \multicolumn{3}{c|}{$g_1'$}\\
\hline
                & LHC 	& LC ($\sqrt{s}= 3$ TeV)& LC ($\sqrt{s}=M_{Z'}+10$ GeV)\\
\hline
 1.0		& 0.0071 & 0.0050               &  0.0026                    \\  
 1.5		& 0.011  & 0.0040               &  0.0032                    \\ 
 2.0		& 0.018  & 0.0028               &  0.0034                          \\ 
 2.5		& 0.028  & 0.0022               &  0.0035                          \\ 
\hline
\end{tabular}
\end{center}
\vskip -0.5cm
\caption{Minimum $g_1'$ value accessible at the LHC and a LC 
for selected $M_{Z'}$ values in our $B-L$ model.
At the LHC we assume $L=100\;{\rm fb}^{-1}$ 
whereas for a LC we take $L=500\;{\rm fb}^{-1}$
at fixed energy and  
$L=10\;{\rm fb}^{-1}$
in energy scanning mode.
\label{mzp-gp-tab}}
\end{table}


In case of the energy scan approach,
when the LC 
energy  is set to $\sqrt{s_{e^+e^-}}=M_{Z'}+10$ GeV 
(assuming $10\mbox{ fb}^{-1}$ of luminosity for each step),
the parameter space can be probed even further
for $M_{Z'}<1.75$~TeV,
as shown in Fig.~\ref{contour2}.
For example, for $M_{Z'}=1$~TeV, 
$g_1'$ couplings can be probed down to the $2.6\times 10^{-3}$,
following a $Z'$ discovery.
Furthermore, one can see
that the parameter space 
corresponding to the mass interval  $500~{\rm GeV}~<M_{Z'}<1$ TeV, 
which the LHC covers better as compared to 
 a LC with fixed energy, can be accessed well beyond the LHC reach 
with a LC in energy scan regime.
Altogether
then, both an ILC,
$\sqrt{s_{e^+e^-}}\leq 1$ TeV) \cite{ILC} and a 
Compact Linear Collider 
(CLIC, $\sqrt{s_{e^+e^-}}\leq 3$ TeV) \cite{CLIC}
design may be able (over suitable regions
of $B-L$ parameter space) to outperform the LHC.


\begin{figure}[htb]
  \subfloat[]{ 
  \label{LCsig-bac1}
  \includegraphics[angle=0,width=0.45\textwidth ]{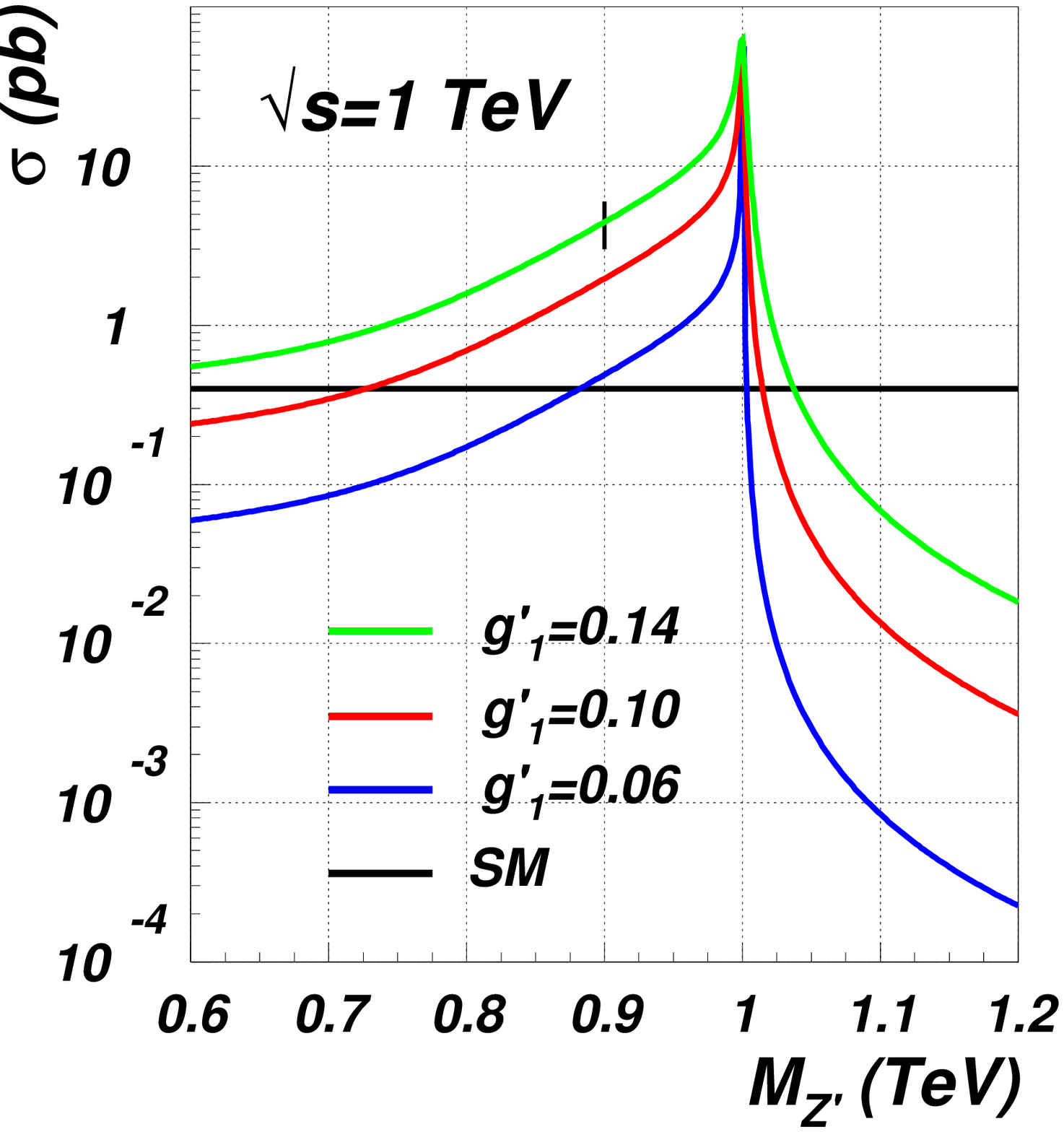}}
  \subfloat[]{
  \label{LCsig-bac2}
  \includegraphics[angle=0,width=0.45\textwidth ]{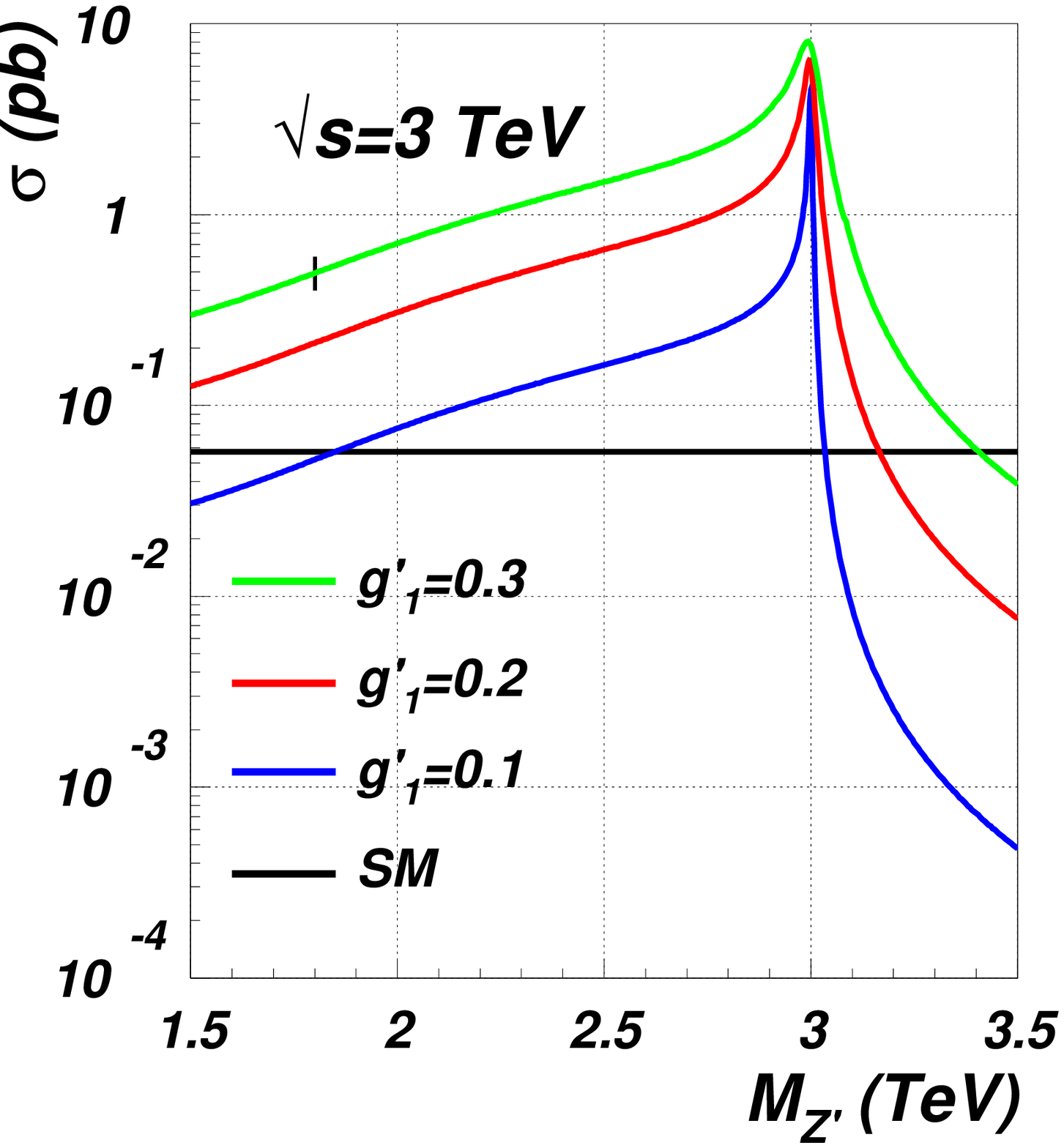}}
  \vspace*{-0.5cm}
  \caption{Cross section for the process $e^+e^-\rightarrow X\to 
\mu^+\mu^-$ for the signal ($X=Z'$) and the SM background ($X=\gamma,Z$, 
independent from $M_{Z'}$) plotted against $M_{Z'}$ at  
(\ref{LCsig-bac1}) a LC with $\sqrt{s_{e^+e^-}}=1$ TeV and 
(\ref{LCsig-bac2}) a
LC with $\sqrt{s_{e^+e^-}}=3$ TeV. (The black vertical bar refers to the mass and coupling combinations excluded
by experimental data, to the left of it.)}
  \label{LCsig-bac}
\end{figure}

Figs.~\ref{LCsig-bac1}--\ref{LCsig-bac2} present the general pattern of
the $Z'$ {production}
cross section {in comparison to} the SM background as a
function of $M_{Z'}$, for two fixed values of $\sqrt{s_{e^+e^-}}$, in
such configurations that the $Z'$ resonance can be either within or
beyond the LC reach for on-shell production. 
The typical enhancement of the signal {at the peak} (now defined as
the $Z'$ sub-channel only) is {either two orders of magnitude above the
background (again defined as $\gamma,Z$ sub-channel only) for
$\sqrt{s}=1$ TeV and $g'_{1}>0.05$ or three orders of magnitude above the
background for $\sqrt{s}=3$ TeV and $g'_{1}>0.1$}. 
{This enhancement} can onset (depending on the value of $g_1'$, hence
of $\Gamma_{Z'}$) several hundreds of GeV before the resonant mass and
falls sharply as soon as the $Z'$ mass exceeds the collider energy.

\begin{figure}[htb]
  \begin{center}
  \subfloat[]{ 
  \label{LClineshape1000}
  \includegraphics[angle=0,width=0.48\textwidth ]{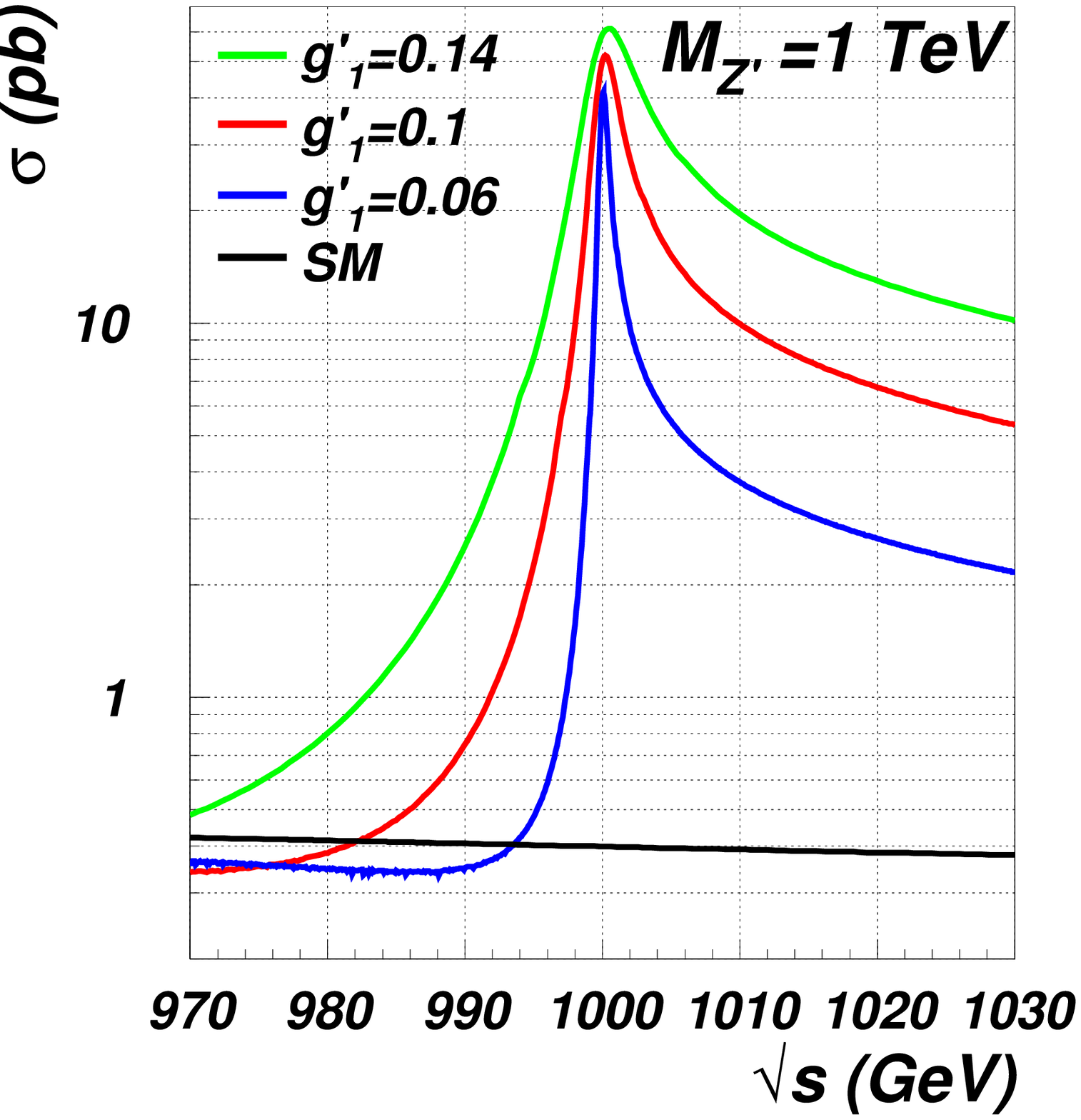}}
  \subfloat[]{
  \label{LClineshape3000}
  \includegraphics[angle=0,width=0.48\textwidth ]{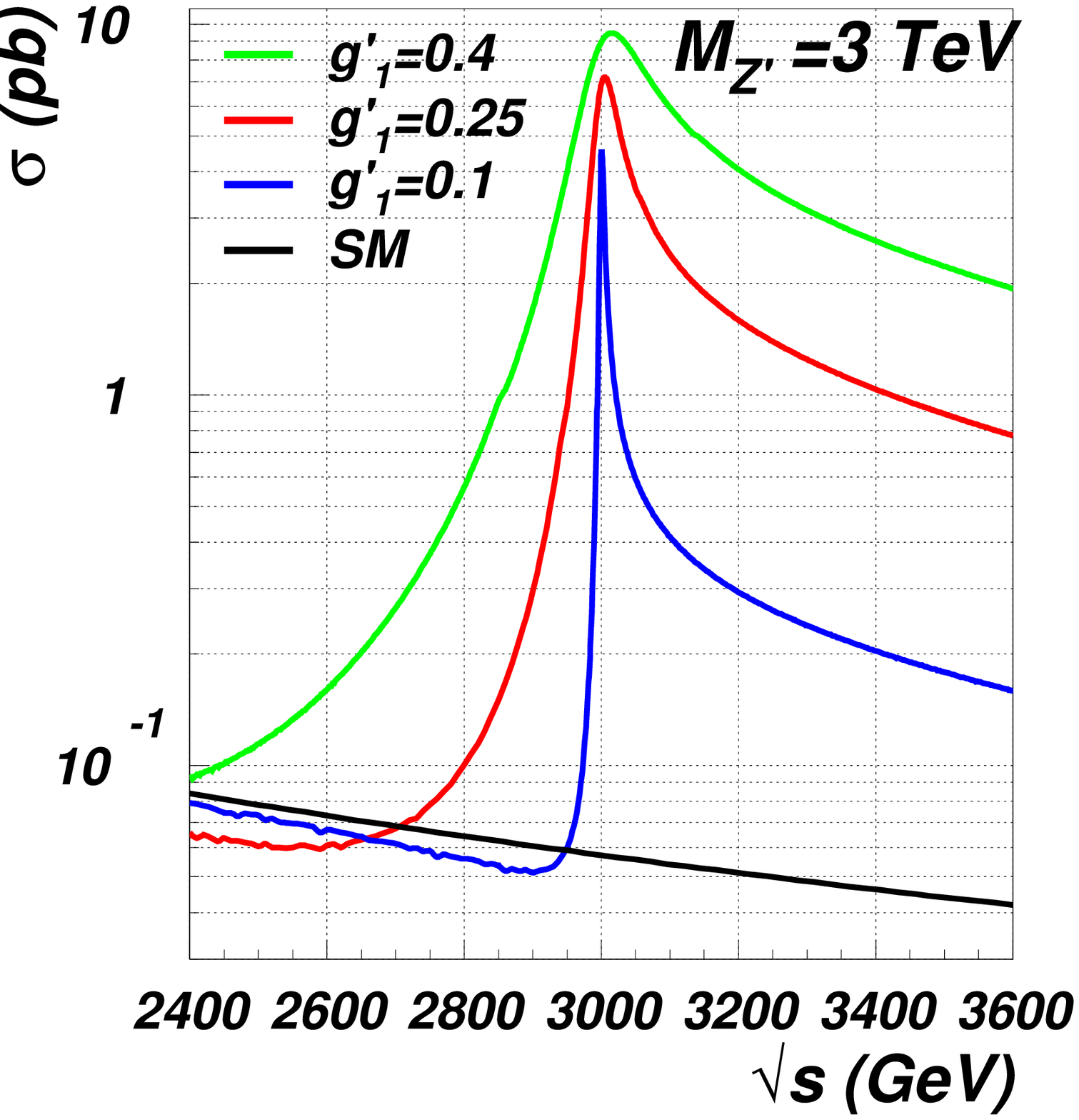}}
  \end{center}
  \vspace*{-0.7cm}
  \caption{Cross section for the process $e^+e^-\rightarrow X\to
\mu^+\mu^-$  cross section for the signal ($X=\gamma,Z,Z'$) and the SM
background ($X=\gamma,Z$) plotted against $\sqrt{s_{e^+e^-}}$ (notice
here the GeV scale) at a LC, for (\ref{LClineshape1000}) fixed $M_{Z'}=1$
TeV and (\ref{LClineshape3000}) fixed $M_{Z'}=3$ TeV.}
  \label{LClineshape}
\end{figure}

\begin{figure}[htb]
  \begin{center}
\includegraphics[angle=0,width=0.8\textwidth ]{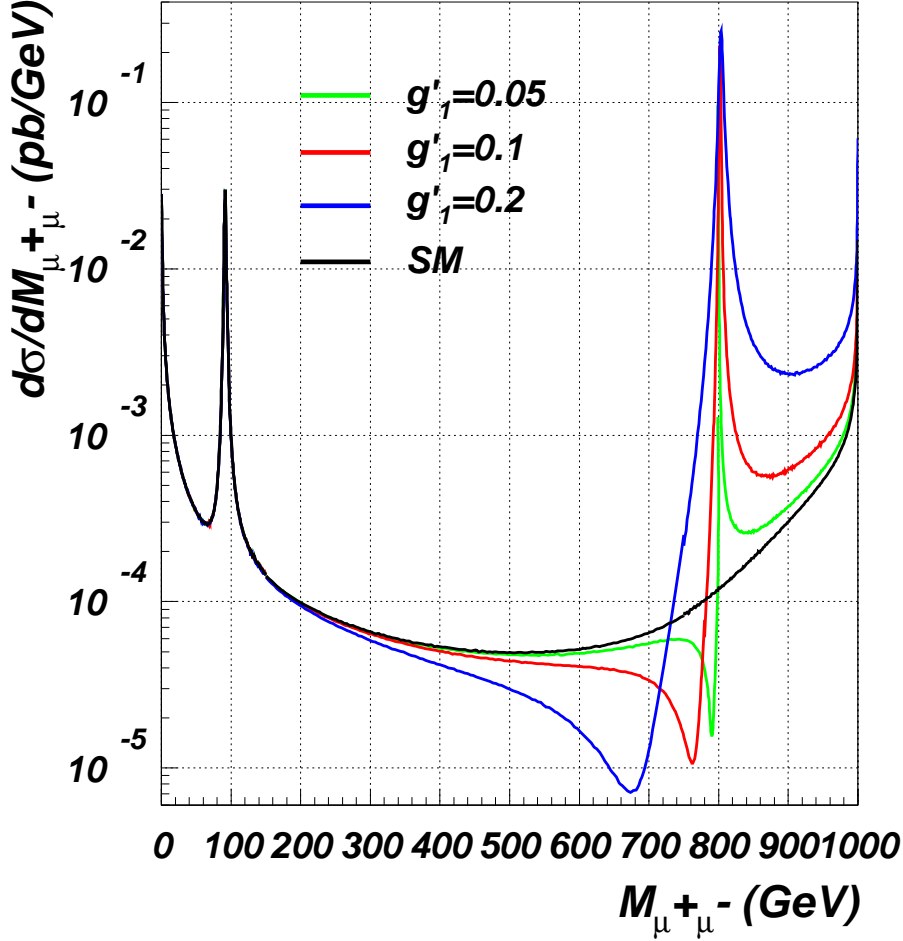}
  \vspace*{-0.7cm}
\caption{$\frac{d\sigma}{dM_{\mu\mu}}(e^+e^-\rightarrow \gamma,Z,Z' \to
\mu^+\mu^-$), for
$\sqrt{s_{e^+e^-}}=1$ TeV, $M_{Z'}=800$ GeV and $g_1'=0.05,~0.1$ and
$0.2$. (Notice that the latter value is shown just for sake of illustration, although already excluded by ref.~\cite{Cacciapaglia:2006pk}, see eq.~\ref{LEP_bound}).}
\label{LCdimuon}
  \end{center}
\end{figure}

Similar effects can be appreciated in
Figs.~\ref{LClineshape1000}--\ref{LClineshape3000}, where the $Z'$ mass
is now held fixed at two values and the LC energy is finely scanned
around the resonance.
In these last two plots, one can neatly appreciate the effects of the
ISR, implying that the maximum cross section (i.e., the one at the $Z'$
peak) is actually achieved for LC energy values 
higher than the $Z'$ mass. Notice that
this energy shift is proportional to the the $Z'$ width 
(i.e., the larger the stronger the $g'_1$ coupling) and is an example
of the radiative return mechanism, whereby ISR effectively modulates
$\sqrt{s_{e^+e^-}}$ over a wide mass range (below the maximum, the
machine energy itself), so that, even at a fixed LC energy, one can
reconstruct the $e^+e^-\to \mu^+\mu^-$ line shape by simply plotting the
di-muon invariant mass distribution, $M_{\mu\mu}$: see
Fig.~\ref{LCdimuon}
(for an illustrative combination of $\sqrt{s_{e^+e^-}}$, $M_{Z'}$ and
$g'_1$'s). 
    
\begin{figure}[htb]
  \begin{center}
  \subfloat[]{ 
\label{LCinter1000}
  \includegraphics[angle=0,width=0.48\textwidth ]{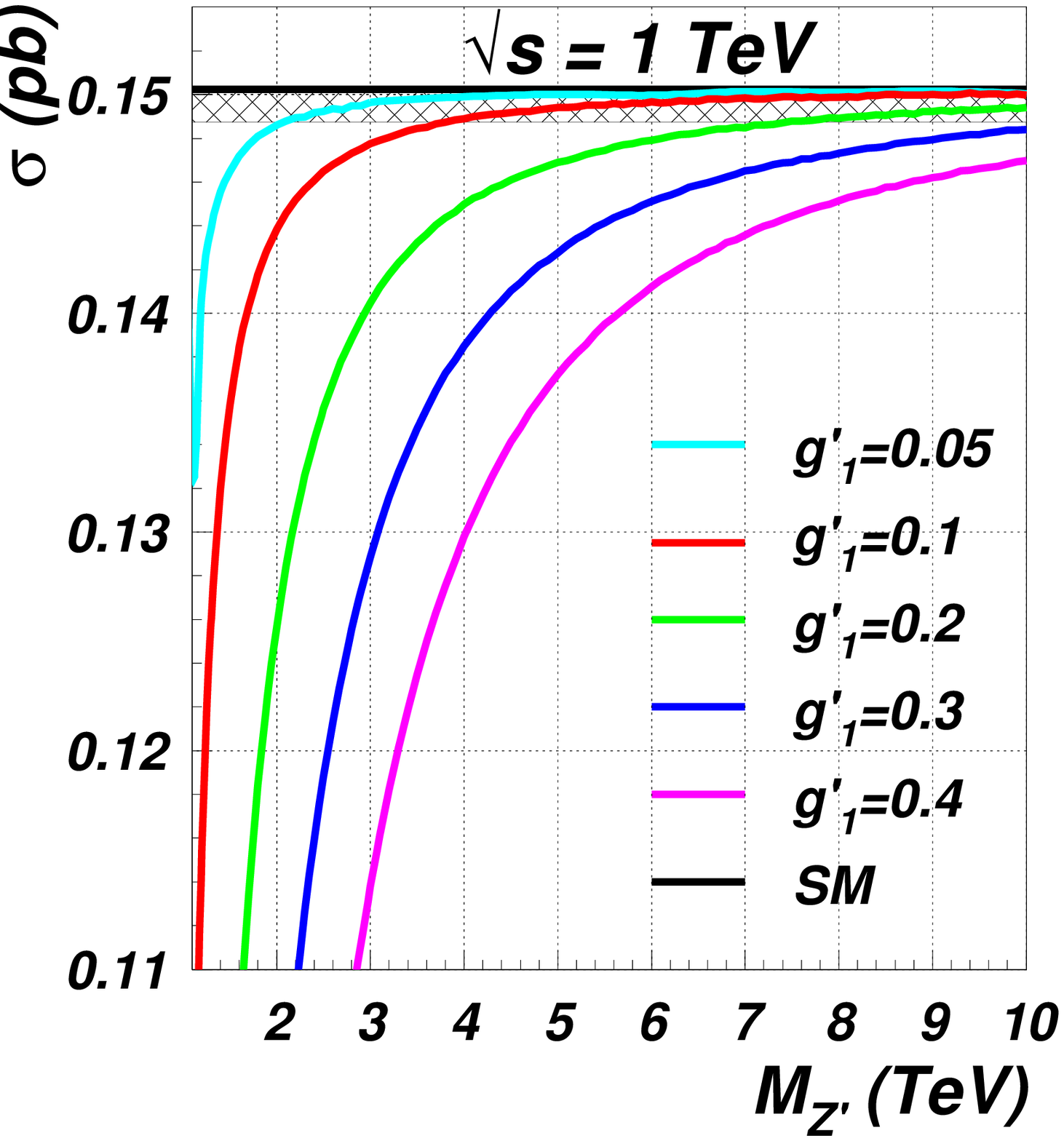}}
  \subfloat[]{
\label{LCinter3000}
  \includegraphics[angle=0,width=0.48\textwidth ]{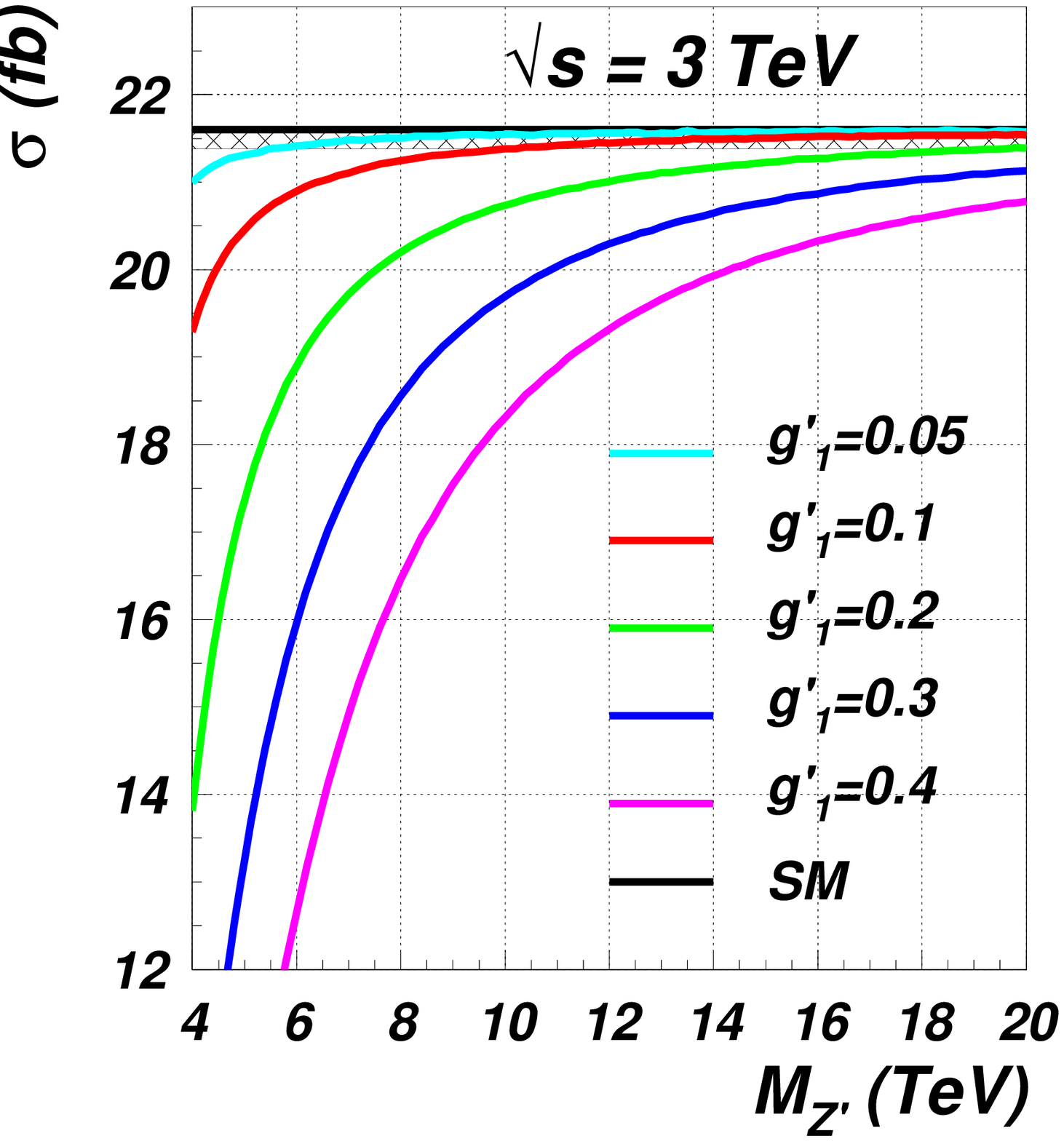}}
  \end{center}
  \vspace*{-0.7cm}
  \caption{Cross section for the process $e^+e^-\rightarrow \gamma,Z,Z' 
\to  \mu^+\mu^-$ plotted against $M_{Z'}$, for (\ref{LCinter1000}) 
$\sqrt{s_{e^+e^-}}=1$ TeV and  (\ref{LCinter3000}) $\sqrt{s_{e^+e^-}}=3$ 
TeV. Notice that we have implemented here the cut $M_{\mu\mu}>200$ GeV.
The shading corresponds to a 1\% 
deviation from the SM hypothesis.}
  \label{LCinter}
\end{figure}

\begin{figure}[htb]
  \begin{center}
  \subfloat[]{ 
\label{LCinterference1000}
  \includegraphics[angle=0,width=0.48\textwidth ]{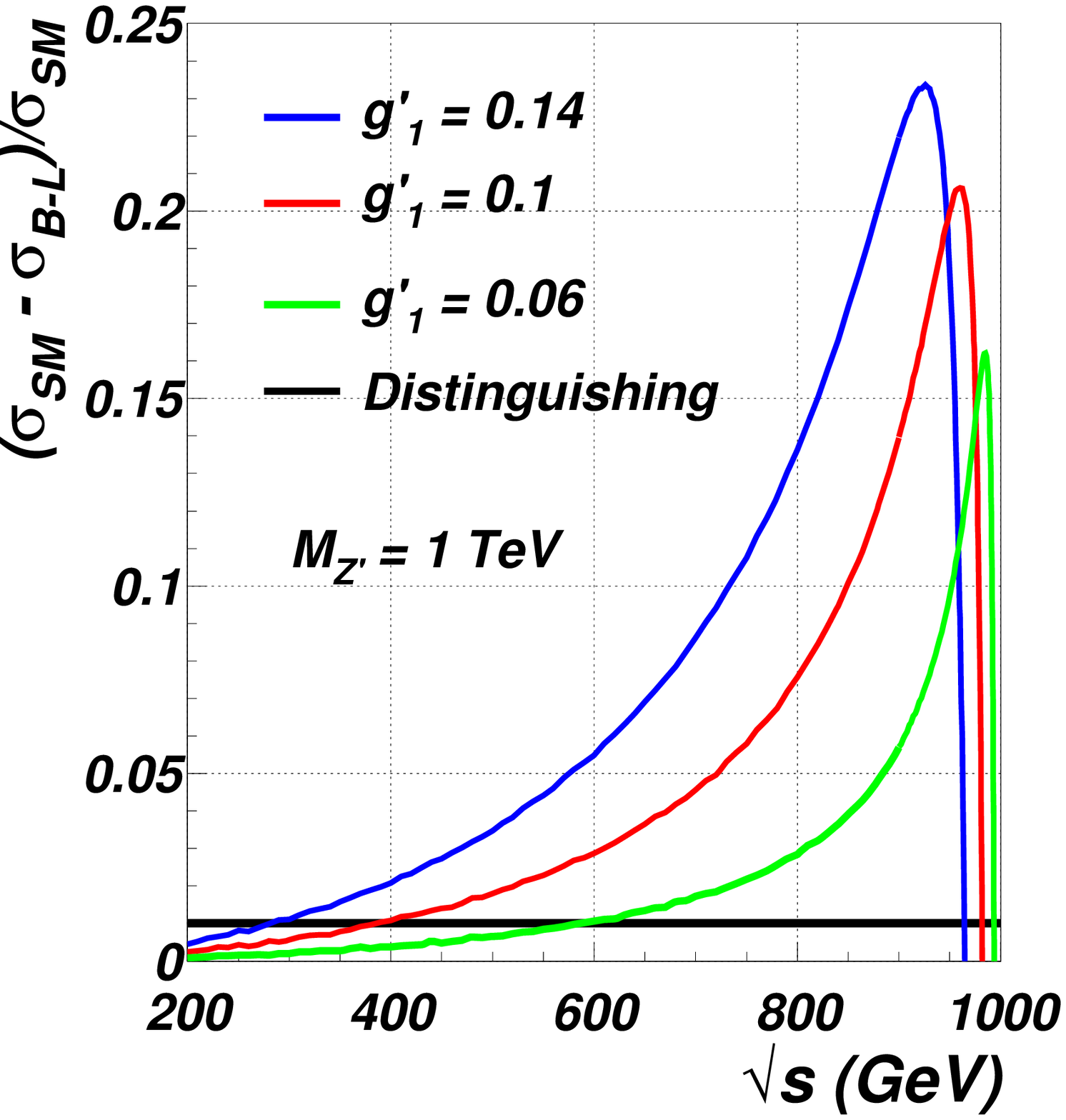}}
  \subfloat[]{
\label{LCinterference3000}
  \includegraphics[angle=0,width=0.48\textwidth ]{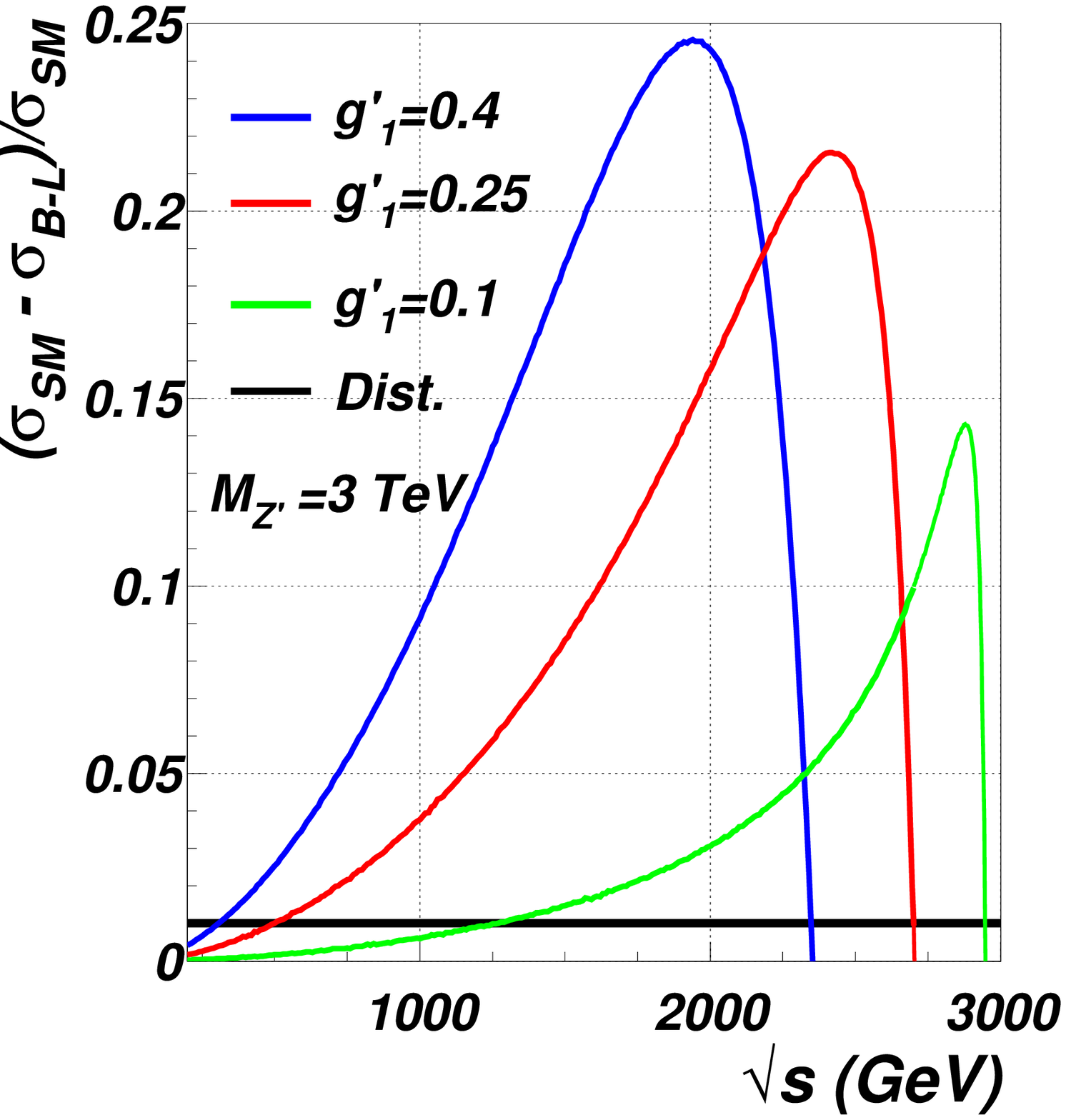}}
  \end{center}
  \vspace*{-0.7cm}
  \caption{
  The relative difference for the cross section of the process  $e^+e^-\rightarrow
\mu^+\mu^-$ between the $B-L$ scenario and the SM  plotted against $\sqrt{s_{e^+e^-}}$,
for (\ref{LCinterference1000}) $M_{Z'}=1$ TeV and 
(\ref{LCinterference3000}) $M_{Z'}=3$ TeV.
The horizontal line corresponds to a 1\% deviation from the SM hypothesis.}
\label{LCinterference}
\end{figure}

\begin{figure}
\begin{center}
\includegraphics[angle=0,width=0.8\textwidth ]{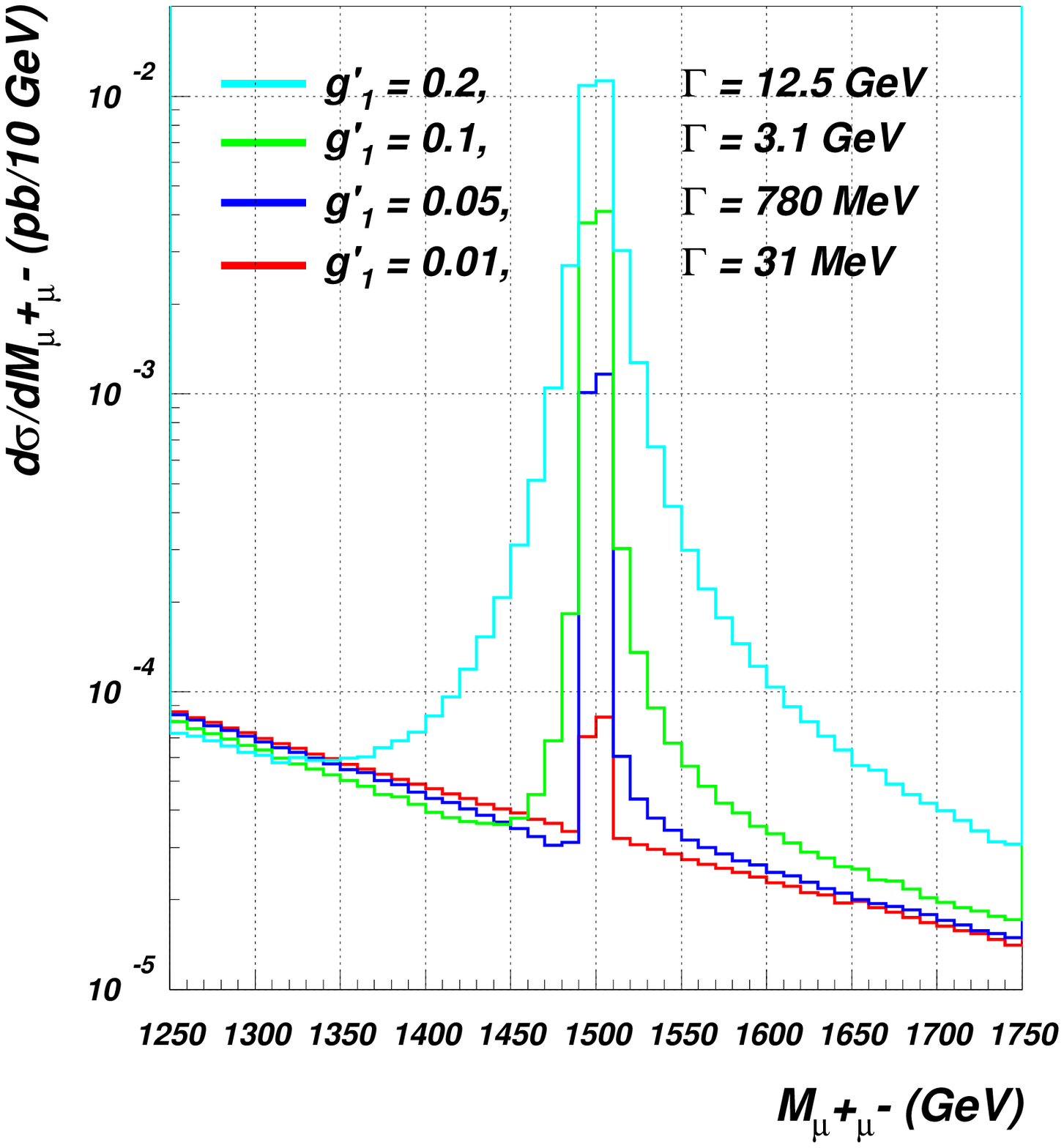}
  \vspace*{-0.7cm}
\caption{$\frac{d\sigma}{dM_{\mu \mu}} (pp\rightarrow \mu^+\mu^-)$ in the
$B-L$ model at the LHC ($\sqrt{s_{pp}}=14$ TeV), with $M_{Z'}=1.5$
TeV, using a 10 GeV binning.}
\label{LHCinterference}
\end{center}
\end{figure}

While the potential of future LCs in detecting $Z'$ bosons of the $B-L$
model is well established whenever $\sqrt{s_{e^+e^-}} \geq M_{Z'}$, we
would like to remark here upon the fact that, even when
$\sqrt{s_{e^+e^-}} < M_{Z'}$,  there is considerable scope to establish
the presence of the additional gauge boson, through the interference
effects that do arise between the $Z'$ and SM sub-processes ($Z$ and
photon exchange). 
Even when the $Z'$ resonance is beyond the kinematic reach of the
LC, significant deviations are nonetheless  visible in the di-muon  line
shape of the $B-L$ scenario considered, with respect to the the SM case.
This is well illustrated in
Figs.~\ref{LCinter1000}--\ref{LCinter3000} for the case of  $\sqrt{s_{e^+e^-}}$
held fixed and $M_{Z'}$ variable (in terms of absolute rates) and in
Figs.~\ref{LCinterference1000}--\ref{LCinterference3000} for the case of 
$M_{Z'}$ held fixed and $\sqrt{s_{e^+e^-}}$ variable (in terms of
relative rates).
{Notice that in 
the studies presented in Figs.~\ref{LCinter1000}--\ref{LCinter3000} 
we have applied a useful kinematical cut $M_{\mu\mu}>200$ GeV, aimed at 
eliminating the production of a SM $Z$-boson due to the radiative return
mechanism as well as  enhancing the aforementioned
 interference effects.}
Incidentally, also notice that such strong interference effects do not onset
in the case of the LHC, as it can clearly be seen from Fig.~\ref{LHCinterference},
owing to smearing due to the PDFs\footnote{See 
also Fig.~7 of Ref.~\cite{B-L:LHC}.}.

\begin{figure}
\begin{center}
  \includegraphics[angle=0,width=0.8\textwidth ]{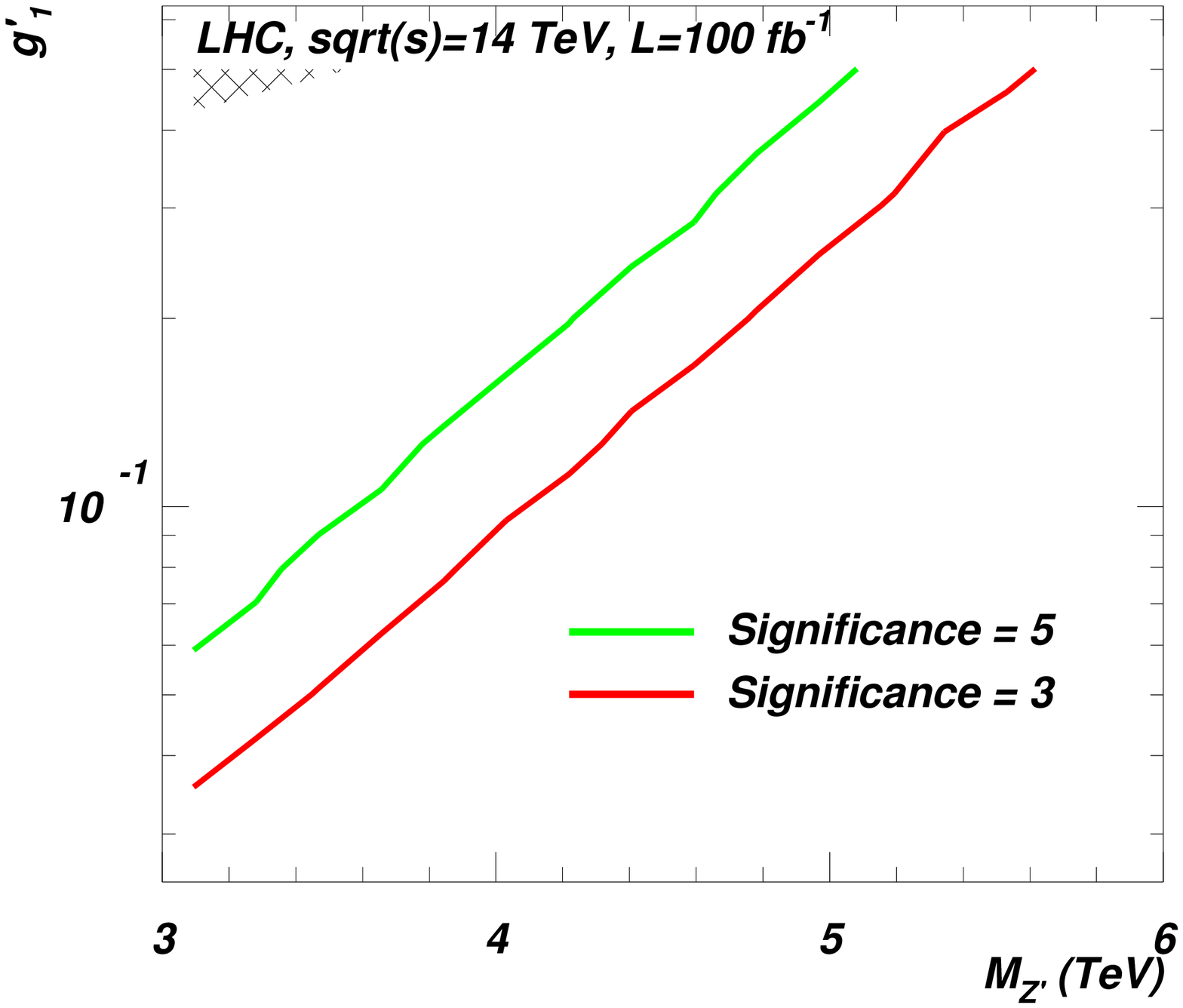}
  \vspace*{-0.7cm}
 \caption{Significance contour levels plotted against $g_1'$ and $M_{Z'}$
at the LHC for $L=100\;{\rm fb}^{-1}$ ($\sqrt{s_{pp}}=14$ TeV, $M_{Z'} \geq 3$ TeV). The
shaded area corresponds to the region of parameter space excluded experimentally, in
accordance with eq.~(\ref{LEP_bound}).} \label{LHC_reach_3TeV}
\end{center}
\end{figure}

{In Figs.~\ref{LCinter1000}--\ref{LCinter3000} and
Figs.~\ref{LCinterference1000}--\ref{LCinterference3000} we have 
assumed and indicated a
$1\%$  uncertainty band on the SM predictions (which is quite conservative).
Under the assumption that SM di-muon production 
will be known with a 1\% accuracy we would like to illustrate
how the LHC $3\sigma$ observation potential of a heavy  $Z'$
(Fig.~\ref{LHC_reach_3TeV})
is comparable to a LC indirect sensitivity to the presence of a $Z'$,
even beyond the kinematic reach of the machine.
This is shown in  Tab.~\ref{mzp-gp-tab_ind}, which 
clearly shows  that a CLIC type LC will be (indirectly) sensitive 
to much heavier $Z'$ bosons than the LHC.
For example, for $g'_1 = 0.1$, such a machine 
would be sensitive to a $Z'$ with mass
up to 10 TeV whilst the LHC can observe a $Z'$ with  mass  below 4 TeV
(for the same coupling).
Even a LC with $\sqrt{s_{e^+e^-}} = 1$ TeV (a typical ILC energy)
will be indirectly sensitive to larger $M_{Z'}$ values that the LHC,
for large enough  values of the $g'$ coupling.
For example, such a machine will be sensitive  to  a $Z'$ with
mass up to $7.5$ TeV for $g'_1 = 0.2$
whilst the  LHC would be able to observe a ${Z'}$ only below $4.7$ TeV or so
(again, for the same coupling).
}

\begin{table}[htb]
\begin{center}
\begin{tabular}{|l|l|l|l|}
\hline
$g_1'$  & \multicolumn{3}{c|}{$M_{Z'}$ (TeV)}\\
\hline
                & LHC ($3\sigma$ observation)	& LC ($\sqrt{s}= 1$ TeV, 1\% level)& LC ($\sqrt{s}= 3$ TeV, $1\%$ level)\\
\hline
 0.05		& 3.4  & 2.2               &  5.5                   \\
 0.1		& 4.1  & 3.8               &  10                    \\  
 0.2		& 4.7  & 7.5               &  19.5                  \\ 
\hline
\end{tabular}
\end{center}
\vskip -0.5cm
\caption{Maximum $M_{Z'}$ value accessible at the LHC and a LC 
for selected $g'_1$ values in our $B-L$ model.
At the LHC we assume $L=100\;{\rm fb}^{-1}$.
\label{mzp-gp-tab_ind}}
\end{table}

\begin{figure}[htb]
  \begin{center}
  \includegraphics[angle=0,width=0.8\textwidth ]{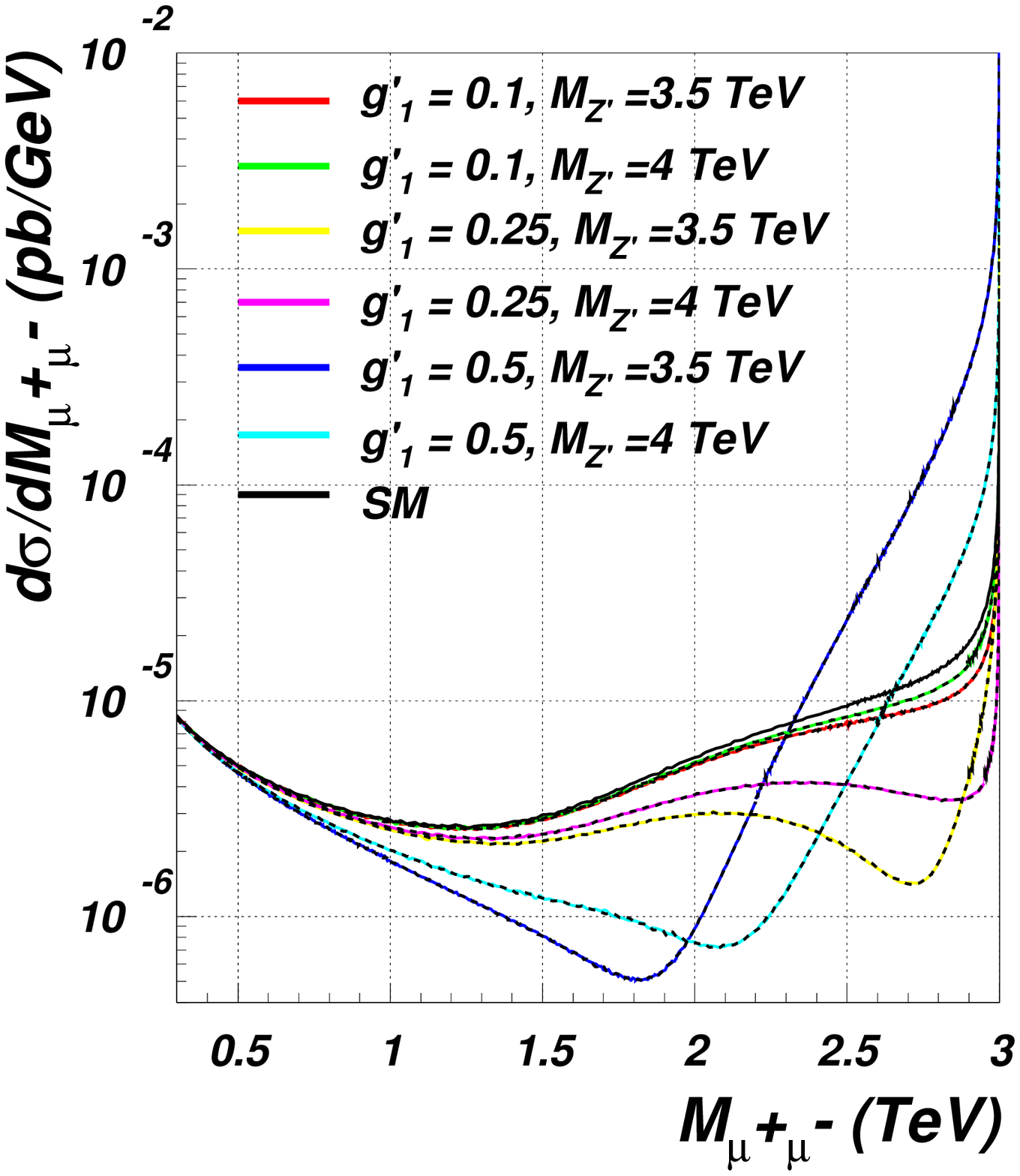}
  \end{center}
  \vspace*{-0.7cm}
  \caption{$\frac{d\sigma}{dM_{\mu\mu}}(e^+e^-\rightarrow 
\mu^+\mu^-)$ in the $B-L$ model, for several combinations of $M_{Z'}$ and $g_1'$, 
treating  $\Gamma_{Z'}$ as an independent parameter: $10$ GeV for colored solid lines, $100$ GeV for black dashed ones.
Here, $\sqrt{s_{e^+e^-}}=3$ TeV.  }
   \label{LCgammaTOT}
\end{figure}

One interesting possibility opened up by such a 
strong dependence of the $e^+e^-\to \mu^+\mu^-$ process in the $B-L$ 
scenario on interferences (up to a 25\% effect judging from, e.g., Fig.~\ref{LCinterference}) 
is to see whether this potentially gives unique and
direct access to measuring the $g'_1$ coupling. In fact, notice that in the case
of $Z'$ studies on or near the resonance (i.e., when $\sqrt{s_{e^+e^-}}\approx M_{Z'}$), 
the $B-L$ rates are strongly dependent on $\Gamma_{Z'}$ (hence on all couplings
entering any possible $Z'$ channel, that is, not only $\mu^+\mu^-$). Instead, when
$\sqrt{s_{e^+e^-}}\ll M_{Z'}$ and  $|\sqrt{s_{e^+e^-}}- M_{Z'}|\gg \Gamma_{Z'}$, one may
expect that the role of the $Z'$ width in such interference effects is minor, the latter
being mainly driven by the strength of $g'_1$. We prove this to be the case in 
Fig.~\ref{LCgammaTOT}, where we have artificially varied the $Z'$ width by a factor
of $10$ in each set of $M_{Z'}$ and $g'_1$ values chosen: the dashed line (corresponding to $\Gamma _{Z'}=100$  GeV) always coincides with the solid one (corresponding to $\Gamma _{Z'}=10$  GeV). Therefore, it is
clear that the dependence on $\Gamma_{Z'}$ is negligible (the more so the larger the difference
$ |\sqrt{s_{e^+e^-}}- M_{Z'}|$) whereas the one on either $M_{Z'}$ or $g'_1$ is always significant.
Hence, in presence of a known value for $M_{Z'}$ (e.g., from a LHC analysis), one could extract
$g_1'$ from a fit to the line shape. In fact, the same method, to access this coupling,
 could be exploited at future LCs independently
of LHC inputs, as interference effects of the same size also appear when $ \sqrt{s_{e^+e^-}} > M_{Z'}$: see again
Fig.~\ref{LCdimuon}.

\section{Conclusions}
\label{Sec:Conclusions}
In summary, we have {demonstrated
the unique potential of future $e^+e^-$
LCs in discovering $Z'$ bosons produced resonantly  
via the $e^+e^-\to\mu^+\mu^- $ process
within  the minimal $U(1)_{B-L}$ extension of the SM.
The scope in this respect of future LCs
operating in the TeV range can be well beyond the reach of the LHC,
in line with what had already been assessed in the literature 
concerning generic $Z'$ scenarios.

We have also presented  the  indirect sensitivity of LCs
to a $Z'$ below its production threshold, assuming a 1\% combined uncertainty
on the $e^+e^-\to\mu^+\mu^-$ production cross section.
For example, for $\sqrt{s_{e^+e^-}}=1(3)$ TeV, one can access $Z'$ masses
up to 2.2(5.5) TeV for $g_1'=0.05$. 
If the value of this coupling is four times larger, an ILC(CLIC) setup 
would be respectively sensitive to the range $M_{Z'}\leq$ 10(20) TeV.
}

Furthermore, in either kinematic configuration
{(i.e, for LCs with centre-of-mass energy below or above the $Z'$ mass)}, 
it may be possible to access both the mass and (leptonic) couplings of
the $Z'$, thereby constraining the underlying model, in parameter space regions  allowed by experimental contraints (see Sect.~\ref{Exp_lim}). 

These results have been obtained by exploiting parton level analyses
 based on exact matrix element calculations appropriately
 accounting for the finite width and all interference effects in the
 $e^+e^-\to \mu^+\mu^-$ channel. We have also taken into account
 beam-shtrahlung effects as well as general detector acceptance
 geometry. Finally, we would like to notice that, {even if our
 model can be fully determined by a direct detection and a line
 shape analysis of the $Z'$ resonance, in case of model checking or
 indirect observation throughout interference effects, the need of
 additional studies could arise. In this connection,} there is
 further room to explore the LC potential to study $Z'$ physics by
 exploiting beam polarisation and/or asymmetries in the cross section,
 which will be reported on separately \cite{preparation}.


\section*{Acknowledgements} 
LB and GMP thank Ian Tomalin for helpful discussions.
AB thanks Andrei Nomerotsky and Tom\'a\u{s} {La\u{s}tovi\u{c}ka} for useful
discussions. SM is financially supported in part by 
the scheme `Visiting Professor - Azione D - Atto Integrativo tra la 
Regione Piemonte e gli Atenei Piemontesi'.



\begin{thebibliography}{99}

\bibitem{Buchmuller:1991ce}
  W.~Buchmuller, C.~Greub and P.~Minkowski,
  Phys.\ Lett.\  B {\bf 267} (1991)  395.
\bibitem{see-saw}  
       P. Minkowski, 
       Phys. Lett. B {\bf 67} (1977) 421; M. Gell-Mann, P. Ramond and R. Slansky, in \emph{Supergravity}, eds.
        P. Van Nieuwenhuizen and D. Freedman (North-Holland, Amsterdam, $1979$), p.~$315$; T. Yanagida, in \emph{Proceedings of the 
Workshop
        on the Unified Theory and the Baryon Number in the Universe}, eds. O. Sawadaand and A. Sugamoto (KEK, Tsukuba, $1979$), p.~$95$;
        S.L. Glashow, in \emph{Quarks and Leptons}, eds. M.L\`evy {\it et al.} (Plenum, New York $1980$), p.~$707$;
        R.N. Mohapatra and G. Senjanovi\'c, Phys. Rev. Lett. \textbf{44} ($1980$) $912$.

\bibitem{Fukugita:1986hr}
  M.~Fukugita and T.~Yanagida,
  Phys.\ Lett.\  B {\bf 174} (1986) 45.


\bibitem{B-L:LHC}  L.~Basso, A.~Belyaev, S.~Moretti and C.~H.~She\-pherd-The\-mi\-sto\-cleous,
  arXiv: 0812.4313 [hep-ph].

\bibitem{B-L:rev}  
K.~Huitu, S.~Khalil, H.~Okada and S.K.~Rai,
  Phys.\ Rev.\ Lett.\  {\bf 101} (2008) 181802;
  W.~Emam and S.~Khalil,
  Eur.\ Phys.\ J.\  C {\bf 522} (2007) 625.

\bibitem{LCs}
K.~Abe {\it et al.}, [The ACFA Linear Collider Working Group],
{arXiv:hep-ph/0109166};
T.~Abe {\it et al.}, [The American Linear Collider Working Group],
{arXiv:hep-ex/0106055}; {arXiv:hep-ex/0106056}; {arXiv:hep-ex/0106057};
{arXiv:hep-ex/0106058};
E.~Accomando {\it et al.}  [ECFA/DESY LC Physics Working Group],
  Phys.\ Rept.\  {\bf 299} (1998) 1;
  J.A. Aguilar-Saavedra {\it et al.}, [The 
ECFA/DESY LC Physics Working Group],  {arXiv:hep-ph/0106315};
K.~Ackermann {\it et al.},
preprint DESY-PROC-2004-01, DESY-04-123, DESY-04-123G.


\bibitem{rizzo}
  T.~G.~Rizzo,
{\it In the Proceedings of 1996 DPF / DPB Summer Study on New Directions for High-Energy Physics (Snowmass 96), Snowmass, Colorado, 25 Jun - 12
Jul 1996, pp NEW136}
  [arXiv:hep-ph/9612440].


\bibitem{LCZZ} See, e.g.:
J. Hewett and T. G. Rizzo, Phys. Rept. {\bf 183} (1989) 193 (and 
references
therein); A. Leike, Phys. Rept. {\bf 317} (1999) 143
(and references therein); 
M. Cvetic and P. Langacker, hep-ph/9707451;
A. Djouadi, A. Leike, T. Riemann, D. Schaile and C. Verzegnassi, Z. Phys. 
C {\bf 56}
(1992) 289;
V. Barger et al., Phys. Rev. D {\bf 33} (1986) 1912; T.G. Rizzo, Phys. 
Rev. 
D {\bf 34} (1986)
1438; F. del Aguila, E. Laermann and P.M. Zerwas, Nucl. Phys. B {\bf 297} 
(1988) 
1;
W. Buchmuller and C. Greub, Nucl. Phys. B {\bf 363} (1991) 345.

\bibitem{ILC_RDR}
  J.~Brau {\it et al.}  [ILC Collaboration],
  arXiv:0712.1950 [physics.acc-ph].


\bibitem{calchep}  A.~Pukhov, 
arXiv:hep-ph/0412191.

\bibitem{lanhep}  A.V.~Semenov, 
arXiv:hep-ph/9608488.

\bibitem{calchep_man} See: http://www.ifh.de/$\sim$pukhov/calchep.html. 

\bibitem{ISR} S. Jadach and B. Ward, Comp. Phys. Commun. {\bf 56}
(1990) 351; S. Jadach and M. Skrzypek, Z. Phys. C {\bf 49} (1991) 577.

\bibitem{AguilarSaavedra:2005pw}
  J.~A.~Aguilar-Saavedra {\it et al.},
  Eur.\ Phys.\ J.\  C {\bf 46} (2006) 43.

\bibitem{Weiglein:2004hn}
  G.~Weiglein {\it et al.}  [LHC/LC Study Group],
  Phys.\ Rept.\  {\bf 426} (2006) 47.

\bibitem{CMSdet} 
  G.~L.~Bayatian {\it et al.}  [CMS Collaboration],
preprint CERN-LHCC-2006-001, CMS-TDR-008-1.

\bibitem{ILCdet} 
  T.~Behnke {\it et al.}  [ILC Collaboration],
  arXiv:0712.2356 [physics.ins-det].

\bibitem{Bityukov}
	S.~I.~Bityukov and N.~V.~Krasnikov, Nucl. Instr. and Meth. A{\bf 452} 	(2000) 518.

\bibitem{CTEQ} See: http://durpdg.dur.ac.uk/hepdata/pdf.html.

\bibitem{preparation} L. Basso, A. Belyaev, S. Moretti  and G. M. 
Pruna, work in progress. 

\bibitem{Carena} M. Carena, A. Daleo, B.A. Dobrescu and T.M.P. Tait, 
 Phys. Rev. D {\bf 70} (2004) 093009. 

\bibitem{Cacciapaglia:2006pk}
  G.~Cacciapaglia, C.~Csaki, G.~Marandella and A.~Strumia,
  Phys.\ Rev.\  D {\bf 74} (2006) 033011.

\bibitem{Tevatron_2.3fb}
  T.~Aaltonen {\it et al.}  [CDF Collaboration],
  Phys.\ Rev.\ Lett.\  {\bf 102}, 091805 (2009)

\bibitem{ILC} A.~Djouadi, J.~Lykken, K.~Monig, Y.~Okada, M.~J.~Oreglia
  and S.~Yamashita,
  arXiv:0709.1893 [hep-ph].

\bibitem{CLIC}
G. Guignard (editor) [The CLIC Study Team], preprint CERN-2000-008 (2000).

\end{thebibliography}
\end{document}